\documentclass[a4paper,twocolumn,11pt,accepted=2025-09-28]{quantumarticle}
\pdfoutput=1
\usepackage[utf8]{inputenc}
\usepackage[english]{babel}
\usepackage[T1]{fontenc}
\usepackage{amsmath}
\usepackage{hyperref}

\usepackage{lipsum}
\usepackage{float}
\usepackage{tcolorbox}
\usepackage{xcolor}
\colorlet{dred}{red!80!black}
\usepackage{tabularx}
\usepackage{makecell}
\usepackage{bm,bbm}
\usepackage{subcaption}
\usepackage{caption}
\usepackage{etoolbox}
\usepackage{braket}

\def\parsepdfdatetime#1:#2#3#4#5#6#7#8#9{%
  \def\theyear{#2#3#4#5}%
  \def\themonth{#6#7}%
  \def\theday{#8#9}%
  \parsepdftime
}

\def\parsepdftime#1#2#3#4#5#6#7\endparsepdfdatetime{%
  \def\thehour{#1#2}%
  \def\theminute{#3#4}%
  \def\thesecond{#5#6}%
  \ifstrequal{#7}{Z}
  {%
    \def\thetimezonehour{+00}%
    \def\thetimezoneminute{00}%
  }%
  {%
    \parsepdftimezone#7%
  }%
}

\def\parsepdftimezone#1'#2'{%
  \def\thetimezonehour{#1}%
  \def\thetimezoneminute{#2}%
}

\newcommand*{\thetimezone}{\thetimezonehour:\thetimezoneminute}

\usepackage{amssymb}
\usepackage{appendix}

\usepackage{multirow}

\begin{document}

\title{Effectiveness of the syndrome extraction circuit with flag qubits on IBM quantum hardware}

\author{Younghun Kim}
\email{hpoqh@hanyang.ac.kr}
\author{Hansol Kim}
\email{khshk18@hanyang.ac.kr}
\author{Jeongsoo Kang}
\email{jskang1202@hanyang.ac.kr}
\author{Wonjae Choi}
\email{marchenw@hanyang.ac.kr}
\author{Younghun Kwon}
\email{yyhkwon@hanyang.ac.kr}
\affiliation{Department of Applied Physics, Hanyang University(ERICA), Ansan 15588, Republic of Korea}
\maketitle

\begin{abstract}
 Large-scale quantum circuits are required to exploit the advantages of quantum computers. Despite significant advancements in quantum hardware, scalability remains a challenge, with errors accumulating as more qubits and gates are added. To overcome this limitation, quantum error-correction codes have been introduced. Although the success of quantum error correction codes has been demonstrated on superconducting quantum processors \cite{ibm_rep,ibm_bell,google_qec1,google_qec2} and neutral atom-based systems \cite{neutral}, there have been no experimental reports of error suppression using flag qubits on a quantum processor. IBM’s quantum hardware features a non-topological coupling map, and past developments of quantum error correction codes on this platform have primarily explored the use of flag qubits. Here, we report the successful implementation of a syndrome extraction circuit with flag qubits on IBM quantum computers. Moreover, we demonstrate its effectiveness by considering the repetition code as a test code among the quantum error-correcting codes. Even though the data qubit is not adjacent to the syndrome qubit, logical error rates diminish as the distance of the repetition code increases from three to nine. Even when two flag qubits exist between the data and syndrome qubits, the logical error rates decrease as the distance increases similarly. This confirms the successful implementation of the syndrome extraction circuit with flag qubits on the IBM quantum computer.
\end{abstract}


 
{
Quantum devices are inherently error-prone, making quantum error correction essential for reliable quantum computation \cite{stabilizer, gottesman, boundary, topological, anyons, surface, qecmemory}. To achieve fault tolerance, quantum error-correcting codes must be implemented to mitigate errors and ensure accurate outcomes from quantum circuits \cite{FT1,FT2,FT3}. Among the available platforms, IBM's quantum devices stand out due to their scale and accessibility. Notably, the IBM quantum machine with the largest qubit count adopts a heavy-hexagon topology rather than the conventional square lattice. In this work, we focus on one such device, \texttt{ibm\_kyoto}, which is accessible via the cloud. This architecture is designed to optimize the fidelity of physical gate operations by minimizing crosstalk at the expense of qubit connectivity \cite{heavy-hexagon,hardware}. As a result, implementing quantum error-correcting codes on IBM's heavy-hexagon hardware may require the use of flag qubits or similar techniques to accommodate its specific structure \cite{flag,flag2,flag3,flag4,hh-code,magic,strategy}.
} \\
{Previous studies have focused on evaluating quantum error correction codes that do not have flag qubits in real devices such as the Sycamore device from Google, where physical qubits can have more connectivity when compared to the heavy-hexagon structure \cite{google_qec1,google_qec2,neutral,hh-code,magic,d2_qec,d3_bacon,d3_ion,d3_surf}.} Quantum error correction on current IBM machines has been explored to demonstrate logical encodings \cite{ibm_rep,d3_dia,d3_surper,ibm_qec,melbourne} and to observe error suppression by scaling the distance of a logical qubit \cite{ibm_rep,ibm_bell}. However, to the best of our knowledge, no experimental demonstrations using flag qubits have been reported. Hence, in this study, we demonstrate that a syndrome extraction circuit with flag qubits in the repetition code can be successfully implemented on the IBM quantum machine.\\
{Flag qubits were originally introduced to detect error propagation and prevent correlated faults \cite{flag}. However, they also serve a broader purpose by enabling the mapping of quantum error-correcting codes onto non-trivial topologies, such as the heavy-hexagon lattice of the IBM hardware. To assess the utility of flag qubits in this setting, we evaluate their impact on the code distance of a logical qubit. Specifically, we examine the repetition code, a well-known QEC code designed to correct either bit-flip or phase-flip errors. We test the syndrome extraction circuit both with and without flag qubits on \texttt{ibm\_kyoto}, an IBM quantum machine, and measure its effectiveness in suppressing errors. Because flag qubits in heavy-hexagon architectures cannot always be adjacent to syndrome qubits, additional routing is often needed. In our implementation, we introduce flag qubits that interact with the syndrome qubits via indirect connections \cite{ibm_qec2,ibm_qec3}.} \\
Fig. \ref{fig1} illustrates how we detect an error that happens during the syndrome extraction circuits with initially selected physical qubits in the hardware. The error types considered are either bit-flips or phase-flips. When we verify the performance of the code on \texttt{ibm\_kyoto}, the logical error rate of the repetition code containing the flag qubits decreases as the number of data qubits increases. This implies that the error correction can be performed using the syndrome extraction circuit with flag qubits in the repetition code on \texttt{ibm\_kyoto}. More specifically, quantum error correction for bit-flips or phase-flips can be achieved on IBM devices, even when the data and syndrome qubits are not close.
\section*{Repetition code with flag qubits}

\begin{figure}[t]
\centering
\includegraphics[width=\linewidth]{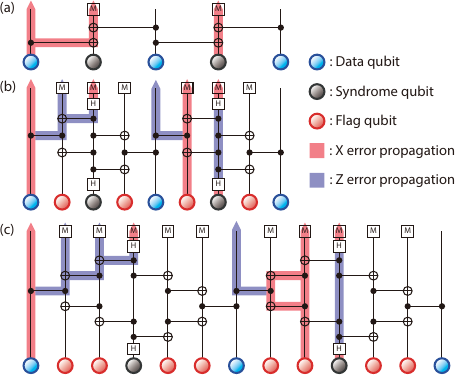}
\caption{Quantum circuit for the Z syndrome extraction circuit: (a) without a flag qubit, (b) with a single-flag qubit, and (c) with double-flag qubits. A syndrome qubit is prepared as $\ket{+}$ by applying the Hadamard gates and a flag qubit is prepared as $\ket{0}$. When a flag qubit exists, there is an indirect interaction between a data qubit and syndrome qubit via CZ and CNOT gates. The blue and red lines display propagated Z and X errors caused by the initialization error on one of physical qubits.}\label{fig1}
\end{figure}

\begin{figure*}[t]
\centering
\includegraphics[width=\linewidth]{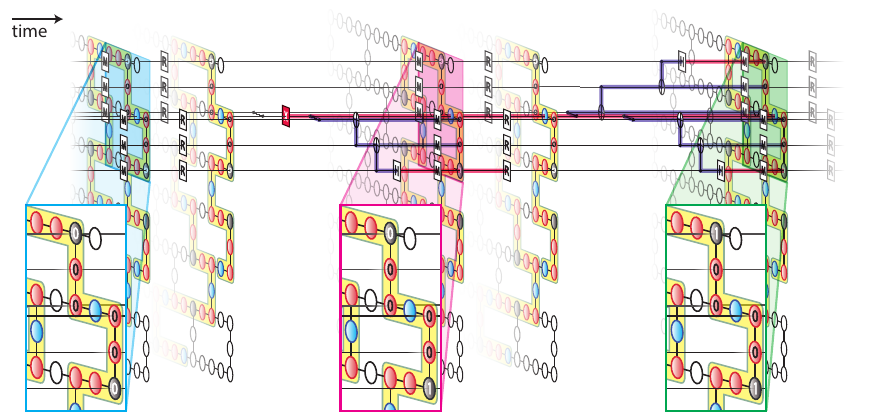}
\caption{Error detection in the repetition code using two flag qubits. The code uses initially selected physical qubits from a heavy-hexagon structure. The code progresses with time represented on the horizontal axis from left to right. The blue, black, and red dots correspond to data, syndrome, and flag qubits, respectively. The code undergoes multiple rounds of a syndrome extraction circuit, which involves reset and measurement gates on each syndrome and flag qubit. When the code uses the Z syndrome extraction circuit, we show the detection of an X error on a data qubit. The error is an example of an ST error. This error disseminates Z or X errors to nearby flag and syndrome qubits, indicated by the blue and red lines, from the data qubit. Over time, outcomes of three syndrome qubits, those closest to the data qubit, are affected: one is from the syndrome extraction circuit round highlighted with the magenta color, where the initial error occurs. The other two are from the subsequent round of the syndrome extraction circuit highlighted with the green color. The error can be detected by comparing consecutive measured outcomes of syndrome and flag qubits.}\label{fig0}
\end{figure*}
A repetition code of distance $d$ consists of a \textit{one-dimensional} array of data qubits of $n_{data} = d $ and syndrome qubits of $ n_{synd} = d-1 $. In a syndrome extraction circuit with flag qubits in the repetition code, a flag qubit is employed to interact with locally separated data and syndrome qubits. This means that the data and syndrome qubits are not adjacent to each other and are connected indirectly by using flag qubits. If a single-flag qubit exists between them, the total number of flag qubits becomes $n_{flag} = 2 n_{synd} $. However, when two flag qubits are used, the structure requires $n_{flag} = 4 n_{synd}$ flag qubits. The repetition code can be denoted by $[n,k,d]_f$, where $n$ and $k$ are the numbers of data qubits and logical qubits, respectively, and $d$ is the distance of the code. Notably, $f=0$ implies no flag qubit; meanwhile, $f=1$ and $f=2$ denote single-flag qubit and double-flag qubits, respectively, between the data and syndrome qubits.\\
We can detect one type of error in a system using the defined stabilizer operators. The stabilizers are either Z or X stabilizers. The Z and X stabilizers are defined with operators from two adjacent data qubits. Therefore, the stabilizer is composed of $i^{th}$ and $({i+1}^{th})$ data qubits' operators, where $i$ belongs to $1,2,...,d-1$. The Z stabilizer can only detect and rectify the X error. Meanwhile, the Z error can be detected and corrected using the X stabilizer. The Z and X stabilizers belong to $S_Z$ and $S_X$, respectively.
\begin{align}
S_Z = \left \langle Z_i Z_{i+1} \right \rangle, S_X = \left \langle X_i X_{i+1} \right \rangle 
\end{align}
 
The logical state of the repetition code can be defined using the stabilizers. The logical states are spanned in a code space $C$.  For example, the code space of the logical states using the Z stabilizers becomes $C = \{ \left| \psi \right\rangle , S_i \left| \psi \right\rangle = + \left| \psi \right \rangle, \left| \psi \right\rangle \in ( \mathbb{C}^2 )^{\otimes d} , S_i \in S_Z \} $. The logical states on the Z basis are denoted as $\ket{0}_L = \ket{0}^{\otimes d}_{data}$ and $\ket{1}_L= \ket{1}^{\otimes d}_{data}$. The logical states on the X basis satisfy $\ket{+}_L = {\frac{1}{\sqrt{2}} } (\ket{0}_L + \ket{1}_L)$ and $\ket{-}_L = {\frac{1}{\sqrt{2}} } (\ket{0}_L - \ket{1}_L)$. However, to check the error correction of X basis logical states, we consider X stabilizers. Hence, $\ket{+}_L $ and $\ket{-}_L$ should be defined as $\ket{+}_L = \ket{+}^{\otimes d}_{data}$ and $\ket{-}_L= \ket{-}^{\otimes d}_{data}$. For both cases, the logical Pauli gates are defined as $Z_L = \prod_{k=1}^{d}{Z_k}$ and $ X_L = \prod_{k=1}^{d}{X_k} $ which change the state of a logical qubit and consist of only the Pauli operators of data qubits.\\
 
If $i^{th}$ stabilizer ($S_i$) anti-commutes with an error of a data qubit ($U_\epsilon$), the error in the data can be detected. 

\begin{align}
S_i U_\epsilon = - U_\epsilon S_i
\end{align}
The error in the data qubit that anti-commutes with the stabilizer can be noticed by performing a subroutine quantum circuit, that is, a syndrome extraction circuit. Therefore, syndrome extraction circuits are used to detect the errors. Fig. \ref{fig1} depicts the Z syndrome extraction circuits in the cases of no-flag qubit ($[3,1,3]_{f=0}$), a single-flag qubit ($[3,1,3]_{f=1}$), and double-flag qubits ($[3,1,3]_{f=2}$) between a data qubit and syndrome qubit. The X syndrome extraction circuits are listed in the supplementary material.\\
 
All data qubits interact with their neighboring syndrome qubits directly or through flag qubits to detect errors. Sequences of quantum operators are implemented to detect errors by measuring both the flag and syndrome qubits. In this section, we focus on single-qubit errors and illustrate their propagation through the syndrome extraction circuit, both with and without flag qubits. In the following section, we will address a more complex scenario in which multiple errors occur. Above all, we explain a case in which there is no flag qubit. Every syndrome qubit is prepared as $\ket{0}$ to construct the Z syndrome extraction circuit depicted in Fig. \ref{fig1} (a). To detect an X error in a data qubit, the CNOT operator is employed, with the data qubit serving as the control qubit and the syndrome qubit as the target qubit. The CNOT operations are performed in parallel to minimize the depth of the circuit. Fig. \ref{fig1} (a) depicts how an error propagates in the circuit when an X error occurs in the data qubit on the left side. The bit-flip error propagates through the CNOT operator. Hence, the error affects its adjacent syndrome qubit.\\
 
Next, we explain the case in which there are flag qubits. When flag qubits exist,  the data and syndrome qubits cannot directly interact. If a single-flag qubit exists, as depicted in Fig. \ref{fig1} (b), the error in the data qubit propagates to its neighbor syndrome qubit through the flag qubit. To detect errors, the structure is constructed as follows: The syndrome and flag qubits are initially prepared as $\ket{0}$. The Hadamard operators are applied to every syndrome qubit to change their states to $\ket{+}$. CNOT gates are implemented to create an entangled state for the syndrome and flag qubits. The CZ operator is applied between the data and the adjacent flag qubit. After applying CNOT gates in reverse order, we measure the syndrome and flag qubits on the Z basis. For the structure with a flag qubit between the data and syndrome qubits, when there is an X error in the data qubit, the error propagates to the flag qubit because the X error causes a Z error in the flag qubit through the CZ operator. The Z error cannot influence the flag qubit because the flag qubit is prepared as $\ket{0}$. Subsequently, the Z error, which propagates to the flag qubit, is transferred to the syndrome qubit through the CNOT operator and flips the syndrome qubit's outcome. Therefore, the X error in the data qubit can be detected by measuring the syndrome and flag qubits. For the structure with double-flag qubits in Fig. \ref{fig1} (c), the data qubit error follows the same process, except being spread through another flag qubit next to the closest syndrome qubit. Fig. \ref{fig0} illustrates how errors on a data qubit can propagate through syndrome extraction circuits in the case of the repetition code using two flag qubits. \\
  
In addition, errors are probable in syndrome or flag qubits. Here, let us consider an X error in initializing a syndrome qubit. The Hadamard gate is applied to this error, and the error transforms into a Z error on the syndrome qubit. Because the syndrome qubit is a control qubit of the CNOT operators ahead of the error, the initial error is not propagated to flag qubits for the remaining gates. Therefore, although the error only produces a bit-flip in the measurement result of the syndrome qubit, it does not affect the quantum state of the data or flag qubits.\\
 However, unlike syndrome qubits, errors in flag qubits may interact directly with data qubits. In Fig. \ref{fig1} (b) and (c), when an X error occurs in the preparation of the initial quantum state of a flag qubit, the error undergoes two-qubit operators, which can be categorized as follows: 1. the CNOT operator between two flag qubits, 2. the CZ operator between a flag and data qubit. The first case produces an additional X error in another flag qubit. The CZ gate is applied to the propagated error. After the CZ gate, the propagated error is canceled when the initial error passes the second CNOT operator. In the second case, when the CZ gate is applied, the X error in the flag qubit propagates a Z error to the data qubit. However, the quantum state of the logical qubit is composed of data qubits on a Z basis and only considers X errors when the Z stabilizers are used. We consider an error as a physical gate that is harmful to the encoded logical qubit state in the repetition code, especially for the quantum memory experiments. Therefore, the Z error in the data qubit may not change the computational state of the logical qubit, for which we are interested in whether the errors get corrected or not.

\begin{figure}[t]
\centering
\includegraphics[width=\linewidth]{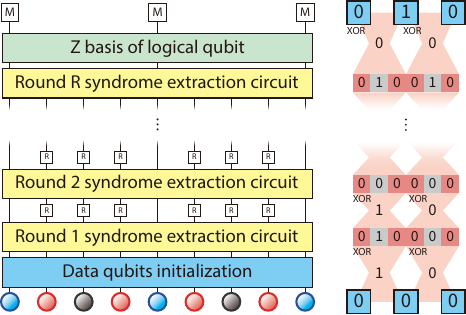}
\caption{ Process example for syndrome extraction round in $[3,1,3]_{f=1}$.
 The figure explains the process of the quantum circuit that samples the result and calculates the syndrome from the measurement result. The quantum circuit consists of three parts: initializing data qubits, R rounds of the syndrome extraction circuit with reset gates on flag and syndrome qubits, and measuring data qubits. The outcomes from each stabilizer extraction circuit are used to obtain the syndrome by employing XOR gates across both temporal and spatial dimensions. }\label{fig2}
\end{figure}

\begin{figure}[t]
\centering
\includegraphics[width=0.7\linewidth]{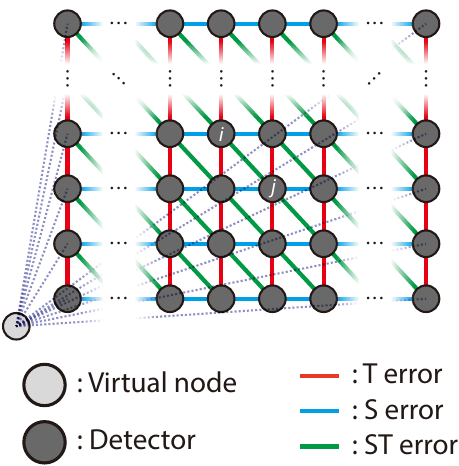}
\caption{ Two-dimensional detector graph that displays a qubit error. Errors that are probable in the system are expressed as the S error (space error), T error (time error), and ST error (space-time error). The virtual node is introduced to show the S error on the boundary data qubits. }\label{fig3}
\end{figure}

 
If the final quantum state of every data qubit can be corrected to its initial quantum state by a correction operator, the quantum correction code can be successful. Fig. \ref{fig2} depicts the quantum memory experiment circuit with the Z syndrome extraction circuits to simulate and evaluate the performance of the code. To perform the single syndrome extraction circuit, the quantum state of every syndrome and flag qubit should be initialized as $\ket{0}$. Hence, after each syndrome extraction circuit round, all the syndrome and flag qubits pass a reset gate. In the initial and final stages, the preparation of the initial state and the measurement of data qubits are performed on the Z basis. Therefore, $\ket{+}_L$ is prepared by preparing $\ket{0}_L$ and applying the Hadamard gate on all data qubits. The measurement at the final state can be obtained by measuring it on the Z basis. When the logical qubit is on the X basis, Hadamard gates are applied to the state before the measurement.

 All measurement results of the syndrome and flag qubits can be listed in a sequence of bits. The binary measurement outcomes from two consecutive rounds pass the XOR gate and create $n_{synd}$ syndrome bits. When syndrome extraction rounds are performed for $N$ times, we can obtain $N-1$ number of syndrome rounds between consecutive measurement rounds, which contains error information. Furthermore, we can build additional syndrome rounds based on the parity between data qubits and the results of the syndrome extraction rounds performed during the first and last times. Therefore, we obtain the results of $N+1$ syndrome rounds and the total $(N+1)\times n_{synd}$ number of bit strings. When there is no error in the system, every syndrome and flag qubit is measured as $\ket{0}$, and all syndrome bits are obtained as 0. Hence, if the syndrome bit is 1, it indicates that an error has been detected, referred to as a detection event.\\
 
Because we prioritize the effectiveness of a syndrome extraction circuit with flag qubits in the repetition code, we use the repetition codes with a distance of $\{ 3, 5, 7, 9 \}$ and execute the syndrome extraction circuit 10 times with or without flag qubits. All quantum circuits are designed and run using the $qiskit$ package \cite{qiskit}. Every case of $\{\ket{0}_L, \ket{1}_L , \ket{+}_L , \ket{-}_L\}$ is considered. In each round, we obtain $5 \times 10^{4}$ number of samples at each logical quantum state, and the total number of samples is $2 \times 10^{5}$. The experiment is conducted during each measurement round and initial state. Gate error rates can vary over time; therefore, we consider the error properties specific to each case for decoding and correcting errors. During the execution of a quantum circuit, dynamic decoupling sequences are added to all the data qubits during its idling period \cite{dd}. Moreover, we use a conditional reset gate to reduce the idling time of the data qubits in the syndrome extraction circuit \cite{ibm_qec3}.

\begin{figure}[t]
\centering
\includegraphics[width=0.7\linewidth]{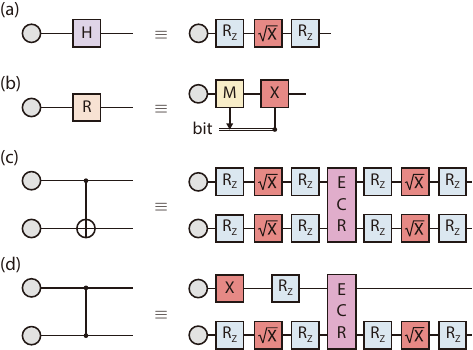}
\caption{ Basis gate decomposition of \texttt{ibm\_kyoto}. \texttt{ibm\_kyoto} uses $\{ R_Z , \sqrt{X}, X, \mathrm{ECR} \}$ gate set as basis gates. The construction of logic gates of repetition code in terms of basis gates of \texttt{ibm\_kyoto} are provided: (a) Hadamard gate, (b) conditional reset gate, (c) CNOT gate, and (d) CZ gate. Detailed information can be found in the supplementary material.}\label{fig4}
\end{figure}

\subsection*{Detector graph}
 
We can infer the most probable correction operator by analyzing the syndromes obtained from the measurement results. A detector graph is constructed for this purpose. Fig. \ref{fig3} depicts a two-dimensional detector graph in which the node denotes a parity of measurement outcome bits that is deterministic when the code is noiseless, in time and space, and the edges represent errors that flip the corresponding detectors \cite{pymatching2}. These errors create a detection event that flips the outcome of a detector to 1. Edges can be classified into T, S, and ST errors. The S(T) error denotes the error in the data qubit (syndrome or flag qubit), which produces a pair of detection events in the corresponding detectors. The ST error denotes the case of consideration of an error that produces two detection events diagonally such as an error of two-qubit gates on a data qubit, which can be seen in Fig. \ref{fig0}. A combination of these three types can express the errors in a system. The virtual node in the graph is employed to denote the error in the data qubit on the boundary (S error). Each edge has a weight calculated based on its error probability.\\

\begin{figure}[t]
\centering
\includegraphics[width=\linewidth]{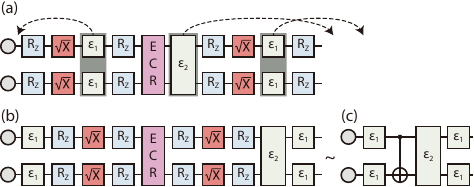}
\caption{ Quantum circuit for the CNOT gate with quantum error channels. $\epsilon_1$ and $\epsilon_2$ denote the depolarizing error channels of single- and two-qubit, respectively. (a) The CNOT gate can be constructed based on the gates of \texttt{ibm\_kyoto} with depolarizing error channels. (b) The errors that occur during gates can be re-expressed as the error channels surrounding the CNOT gate. (c) The noisy CNOT gate can be fragmented into the error channels before and after the ideal CNOT gate.} \label{fig5}
\end{figure}

\begin{figure*}[t]
\centering
\includegraphics[width=\linewidth]{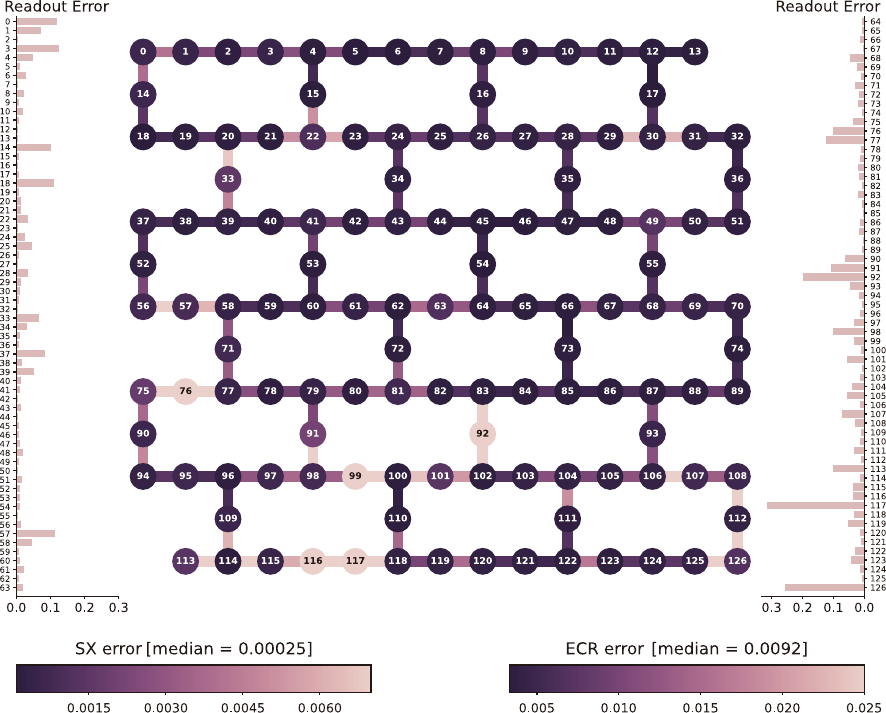}
\caption{ Error rate of single qubit gate, ECR gate, and readout in 127 physical qubits of \texttt{ibm\_kyoto} at 2023-12-25 11:42:34+00:00. While each readout error is displayed on the side, the error rate of the single-qubit and ECR gate are indicated using dots and edges, respectively, in the graph. The median error rate of the single qubit gate (ECR gate) is $2.5\times10^{-4}$ ($9.2\times10^{-3}$)}\label{fig6}
\end{figure*}

When finding the most plausible correction operators, the decoder will use the detector graph for computing the weights of every pair of detection events in the given syndrome bit-string. We decompose complex errors that create more than a pair of detection events into space-, time-, and spacetime-like errors and estimate weights for every pair of edges based on the detector graph. This process will require pairing a set of detection events using decomposed errors in a detector graph. To pair them, conventionally, the blossom algorithm has been widely used; its complexity is $O(s^3 log(s))$, where $s$ is the number of detection events. To reduce it for real-time decoding, recent research has put efforts into balancing the time and accuracy, such as the Pymatching decoder and the sparse blossom algorithm \cite{pymatching,pymatching2}.\\
 Two methods can be used to evaluate the weights of a detector graph. The first method exploits the error rates of every gate in the IBM hardware \cite{strategy}. The second method uses the correlation between the results of the measured samples from the quantum circuit.
\begin{figure*}[t]
\centering
\includegraphics[width=\linewidth]{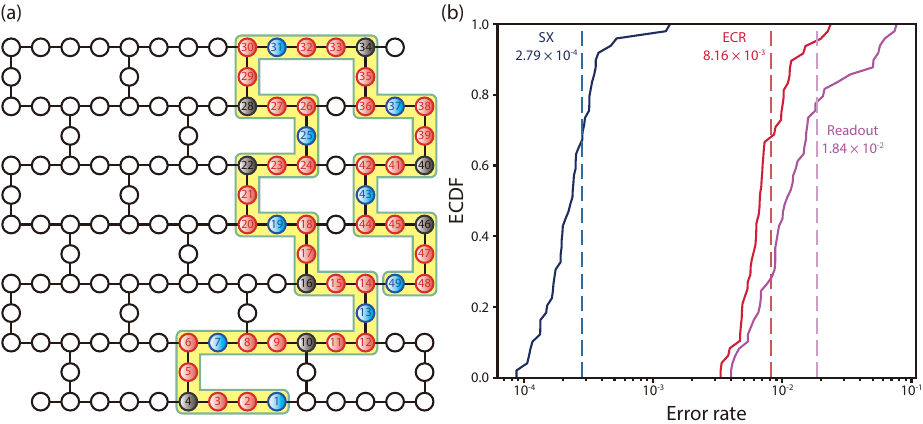}
\caption{The selected 49 physical qubits for $[9,1,9]_{f=2}$ and the error rate of  \texttt{ibm\_kyoto}.
(a) The selected 49 physical qubits for $[9,1,9]_{f=2}$ are shown with color dots with respect to their types: data, syndrome, and flag qubits. The blue dots correspond to data qubits, black and red dots represent syndrome, and flag qubits, respectively. (b) The graph shows the ECDF of gate errors for selected physical qubits at $[9,1,9]_{f=2}$. In this figure, error rates of quantum gates are considered when the initial logical state is $\ket{0}_L$. The cumulative distribution of each error of ECR, SX($\sqrt{X}$), and readout are plotted as a function of the physical error rate. The dot lines correspond to their mean values.}\label{fig7}
\end{figure*}
\subsection*{Hardware-based detector graph}
 
The quantum memory experiment with the syndrome extraction circuit rounds of the repetition code consists of initialization, single-qubit gates, CNOT gates, and measurements. However, the quantum hardware of \texttt{ibm\_kyoto} uses  $\{ \mathrm{ECR}, I, R_Z , \sqrt{X}, X, M \}$ as the basis gate. Notably, \texttt{ibm\_kyoto} uses an echoed cross-resonance gate (ECR) for a two-qubit operator to create entanglement \cite{ecr}. Therefore, quantum circuits should be expressed in terms of basis gates in the hardware. Fig. \ref{fig4} shows that $\{ H, R, \mathrm{CNOT}, \mathrm{CZ} \}$ can be expressed by combining gates in \texttt{ibm\_kyoto}. Although the CNOT gate consists of eight $R_Z$, four $\sqrt{X}$, and a single ECR gate, the CZ gate can be fragmented into eight single qubit gates and an ECR gate. However, based on the parameters used for the $R_Z$ gate, there are various methods to construct two-qubit gates, and a detailed explanation can be found in the supplementary material.\\
 
As explained previously, we can evaluate the weights of a detector graph using the error rates of every gate in the IBM hardware. The quantum circuit constructed with only Clifford gates can be simulated efficiently \cite{clifford1,clifford2}. Hence, an equivalent quantum circuit needs to contain a gate set with $\{ \mathrm{CNOT}, \mathrm{CZ}, I, X, Z, H, M, R \}$ and an error channel with Pauli gates to obtain the weight of a detector graph \cite{stim}. Fig. \ref{fig5} depicts the case for the noisy CNOT gate when there are quantum error channels such as $\epsilon_1$ and $\epsilon_2$, which denote the depolarizing channels of the single-qubit and two-qubit gates. The error rate for both the single- and two-qubit gates can be adopted as the error rate from the corresponding gate in a real device. Because the depolarizing error channel and unitary gates are commutative, the noisy gates consist of the corresponding ideal gate and error channels.

{All basis gates, except the \(R_Z\) gate, are susceptible to errors \cite{virtual_Z}. The \(R_Z\) gate is implemented as a virtual gate, meaning it does not involve any physical pulse and therefore does not experience hardware-induced errors. As a result, \(R_Z\) is effectively noise-free. In contrast, gates such as ECR, \(\sqrt{X}\), and \(X\) are followed by depolarising noise channels.}
However, the readout and reset gate errors are considered with an X error instead of the depolarizing channel.  While the noisy version for the other gates $\{ \mathrm{CZ}, I, X, H \}$ can also be constructed using depolarizing channels.  We can design a quantum circuit by surrounding the ideal gate with depolarizing error channels, ensuring that the gate consists only of the Clifford gates. Therefore, we can construct a quantum circuit with Clifford gates, simultaneously considering the error rate of the hardware where its basis gates are not Clifford gates. \\
 
To implement a syndrome extraction circuit with flag qubits in the repetition code on \texttt{ibm\_kyoto}, we select a combination of one-dimensional qubits from a heavy-hexagon structure of 127 qubits. Fig. \ref{fig6} depicts the error rates of the SX gate, ECR gate, and measurement in \texttt{ibm\_kyoto} \cite{rd}. The average of the lifetime (T1) and coherence time (T2) are 219.08 $\mu$s and 125.06 $\mu$s, respectively. When a flag qubit is introduced between the data and syndrome qubits, we need 33 qubits in the case of $[9,1,9]_{f=1}$, because it uses 16 flag qubits between the data and syndrome qubits. If double-flag qubits exist between the data and syndrome qubits, we need 49 qubits in the case of $[9,1,9]_{f=2}$, as depicted in Fig. \ref{fig7} (a). Fig. \ref{fig7} (b) shows that the ECR pulse, the measurement, and single qubit gate error rates in the selected qubits for the $[9,1,9]_{f=2}$ are expressed in terms of the empirical cumulative distribution function (ECDF). 

\begin{figure}
\centering
\includegraphics[width=\linewidth]{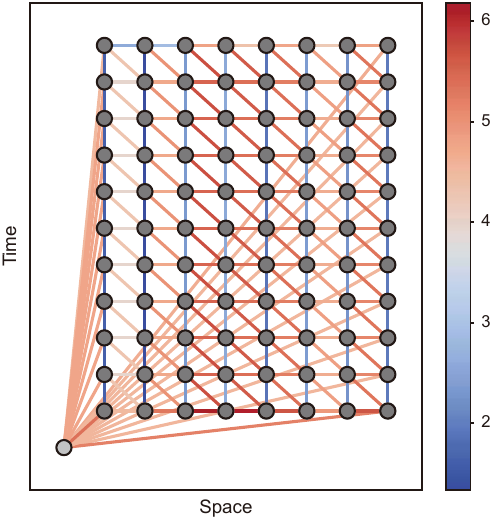}
\caption{ Detector graph for a distance of 9 with double-flag qubits. The horizontal, vertical, and diagonal lines denote the S, T, and ST errors, respectively. The weight of each edge is obtained from the Stim code, which has the depolarizing error channel with the error rates of basis gates based on \texttt{ibm\_kyoto}. }\label{fig8}
\end{figure}

Fig. \ref{fig8} depicts the detector graph of $[9,1,9]_{f=2}$ obtained using the Stim code \cite{stim}. The weight for each edge in the detector graph can be determined by considering the probability of the corresponding error. We assign a lower weight when the probability is high, increasing the likelihood that the MWPM algorithm selects that error as a correction operator. {The edge coloring reflects edge weight: red edges have higher weight and blue edges lower weight respectively.} Further, as the weight decreases, the intensity of the blue color increases. This means that an error corresponding to a lower weight on a blue edge is more likely to occur than others. Therefore, it will be selected more frequently as a correction operator over others with a higher weight.\\ 
\begin{table}[htbp]
  \centering
  \resizebox{1\linewidth}{!}{
  \begin{tabular}{|| c | c | c ||}
    \hline\hline 
    \textbf{Structure $[n,k,d]_f$} & \textbf{Error Type} & \textbf{Avg. weight} \\ [0.5ex] 
    \hline\hline
    \multirow{3}{*}{$[9,1,9]_{f=0}$} & S error & 5.560 \\
    & T error & 4.387 \\
    & ST error & 5.803 \\
    \hline
    \multirow{3}{*}{$[9,1,9]_{f=1}$} & S error & 5.260 \\
    & T error & 2.893 \\
    & ST error & 5.487 \\
    \hline
    \multirow{3}{*}{$[9,1,9]_{f=2}$} & S error & 4.874 \\
    & T error & 2.038 \\
    & ST error & 5.174 \\
    \hline\hline
  \end{tabular}
  }
  \caption{ Average weights of S, T, and ST errors in the hardware-based detector graph when the number of data qubits is nine. }
\end{table}

\begin{figure*}[t]
\centering
\includegraphics[width=\linewidth]{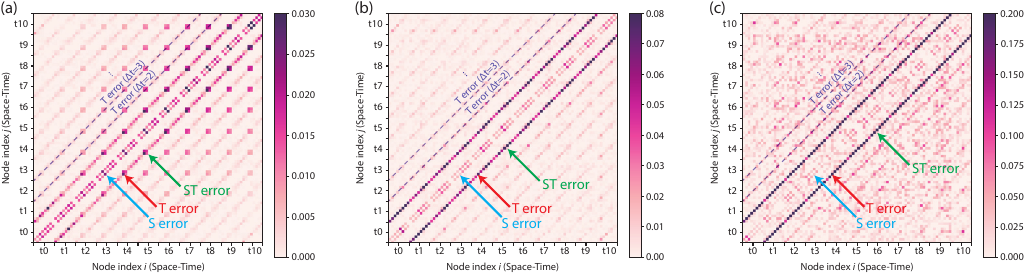}
\caption{Correlation matrices of repetition code at a distance of nine with different numbers of flag qubits. Space-time correlation matrices for (a) $[9,1,9]_{f=0}$, (b) $[9,1,9]_{f=1}$, and (c) $[9,1,9]_{f=2}$ are shown. The matrices are obtained when the initial logical state is $|1\rangle_L$, and the number of rounds in the syndrome extraction is 10. Each pixel can be mapped to the edge of the corresponding detector graph. The S, T, and ST errors are visible along the diagonal line. The dashed diagonal lines correspond to multiple T errors occurring consecutively in a row. The axes represent the labeled detector in the detector graph in the space-time method. Each syndrome round is divided into $t$, which varies from 0 to 10.
}\label{fig9}
\end{figure*}
 
Table 1 lists the average values of the weights for the cases of $[9,1,9]_{f=0}$, $[9,1,9]_{f=1}$, and $[9,1,9]_{f=2}$ with respect to the type of errors: S, T, and ST errors. As listed in Table 1, the weights corresponding to the measurement errors have smaller values regardless of the existence of the flag qubit. This tendency appears in the case of the average error rate of gates of \texttt{ibm\_kyoto}. Furthermore, the weights with flag qubits have lower values than those without flag qubits. This implies that the more flag qubits we use, the more likely the errors are to occur.
\subsection*{Sample-based detector graph}
 
The second method evaluates the statistical correlation between samples and uses it to decode errors \cite{google_qec1,google_qec2}. The main idea of the sample-based detector graph is to utilize statistical correlations between detectors as weights, based on actual syndrome samples. It assumes that every possible error is uncorrelated and independent. We can derive the probability ($p_{ij}$) of detection events on each node of any pair with the given assumption. These probabilities can be obtained through the experimental results, especially using the expectation values of detectors. When $p_{ij}$ is the probability where two detectors of $i$ and $j$ can be simultaneously {flipped},  $p_{ij}$ can be obtained by calculating the probability of detection events on each detector ($\left \langle x_j \right\rangle$, $\left \langle x_i \right\rangle$) or both simultaneously ($\left \langle x_i x_j \right\rangle$). 
\begin{align}
 p_{ij} = {\frac{1}{2}} - {\frac{1}{2}} \sqrt{ 1-  { \frac{ 4( \left \langle x_i x_j \right\rangle - \left \langle x_i \right\rangle \left \langle x_j \right\rangle )}{ 1- 2 \left \langle x_i \right\rangle - 2\left \langle x_j \right\rangle + 4\left \langle x_i x_j \right\rangle} } }
\end{align}
If the error probability $p_{ij}$ has a negative value, it is set to zero.
\begin{figure*}[t]
\centering
\includegraphics[width=\linewidth]{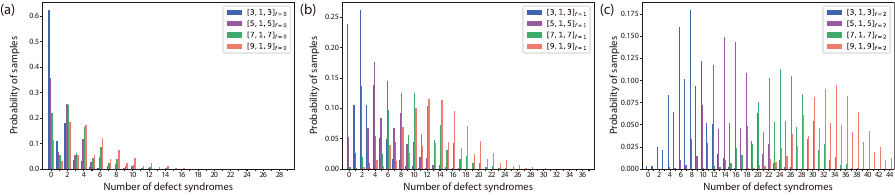}
\caption{ Ratio between the number of corresponding samples and total number of sampled data in terms of the number of detection events for (a) $\lbrack9,1,9\rbrack _{f=0}$, (b) $\lbrack9,1,9\rbrack _{f=1}$, and (c) $\lbrack9,1,9\rbrack _{f=2}$. The data is obtained with 10 rounds of syndrome extraction. The probability is calculated by considering Z basis states $\ket{0}_L$ and $\ket{1}_L$, meaning the total number of samples is $10^{5}$. The number of detection events is plotted when its probability is more significant than $10^{-5}$ for all distances.
    }\label{fig10}
\end{figure*}
The correlation matrix indicates the correlation between detection events. The number of detectors can be considered in terms of the horizontal direction corresponding to space and the vertical direction corresponding to time. The horizontal number is denoted by $s$, which can be $1,2, \dots, d-1$. The vertical number is denoted by $t$, which can be $1, \dots, R+1$, where R is the number of rounds of syndrome extraction. The order of detectors can be determined in two cases: The first is when the priority is space, $N_{s-t}$, which is the space-time type. The second is when the priority is time, $N_{t-s}$, which is the time-space type. Detectors can be labeled based on one of these methods.
\begin{align}
N_{s-t} &= s + (d-1)( t - 1 ) \nonumber \\
N_{t-s} &= t + ( R + 1 )(s-1)
\end{align}
 
Fig. \ref{fig9} (a), (b), and (c) depict the correlation matrix of syndrome samples of $[9,1,9]_{f=0}$, $[9,1,9]_{f=1}$, and $[9,1,9]_{f=2}$, respectively, which are displayed in terms of the forms of the space-time method. These figures exhibit the explicit correlation between detectors in space and time. The probability of detecting two detection events is higher in the presence of flag qubits than in their absence. As we add more flag qubits, we can see a higher probability of an error, which can be broken down into a combination of categorized errors (S, T, and ST error). This implies that the cross-talk effect becomes larger when flag qubits are included. T errors, caused by measurement errors on either syndrome or flag qubits, can occur consecutively over time because they are more frequent than S or ST errors. The phenomenon is observed in all cases, with or without flag qubits, as depicted in Fig. \ref{fig9} (a), (b), and (c) with the dotted diagonal lines.\\
\begin{figure*}[t]
\centering
\includegraphics[width=\linewidth]{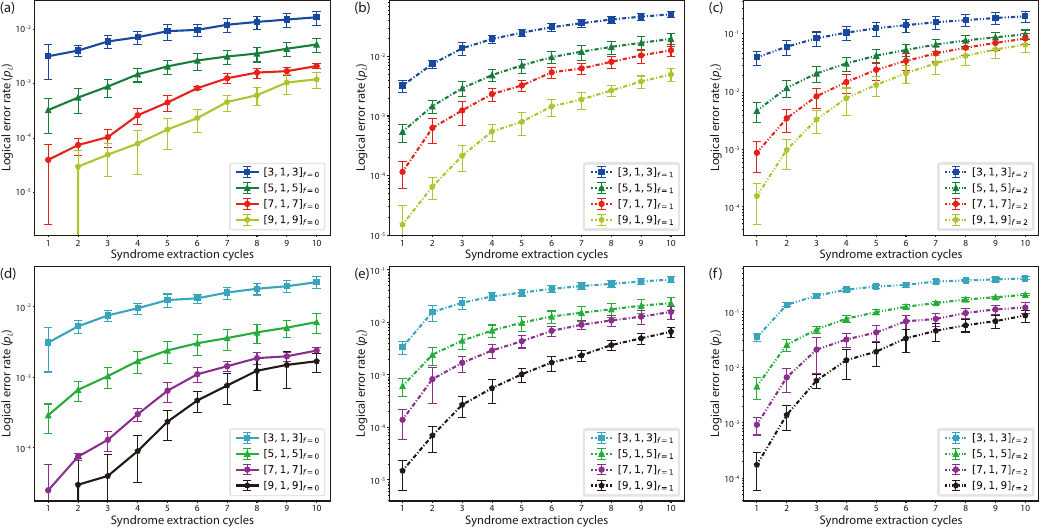}
\caption{Logical error rate of repetition codes. The figure displays the logical error rates of $[d,1,d]_{f=0}$, $[d,1,d]_{f=1}$, and $[d,1,d]_{f=2}$ in terms of syndrome extraction rounds when (a, b, c) hardware-based or (d, e, f) sample-based detector graph is used. The logical error rate is shown as a function of the number of data qubits ($d=3, 5, 7, 9$) with different structures as follows:
    (a, d) The logical error rate of repetition codes with no flag qubit, 
    (b, e) a single-flag qubit, and 
    (c, f) double-flag qubits.
}\label{fig11}
\end{figure*}

\begin{table}[htbp]
  \centering
  \resizebox{1\linewidth}{!}{
  \begin{tabular}{|| c | c | c ||}
    \hline\hline 
    \textbf{Structure $[n,k,d]_f$} & \textbf{Error Type} & \textbf{Avg. weight} \\ [0.5ex]
    \hline\hline
    \multirow{3}{*}{$[9,1,9]_{f=0}$} & S error & 4.416 \\ 
    & T error & 4.350 \\ 
    & ST error & 5.533 \\ 
    \hline
    \multirow{3}{*}{$[9,1,9]_{f=1}$} & S error & 3.868 \\ 
    & T error & 2.680 \\ 
    & ST error & 5.101 \\ 
    \hline
    \multirow{3}{*}{$[9,1,9]_{f=2}$} & S error & 2.800 \\ 
    & T error & 1.120 \\ 
    & ST error & 4.321 \\ 
    \hline\hline
  \end{tabular}
  }
  \caption{Average weights of S, T, and ST errors in the sample-based detector graph when the number of data qubits is nine.}
\end{table}

 
Table 2 lists the average weight of the detector graph obtained from the correlation matrix for the cases $[9,1,9]_{f=0}$, $[9,1,9]_{f=1}$, and $[9,1,9]_{f=2}$. When the weight is derived from the sampled data, time errors are also more likely to occur, similar to the case with the hardware-based detector graph.\\
 
Fig. \ref{fig10} (a), (b), and (c) depict the probability of the samples from the data concerning the number of detection events according to the distance in the cases of no flag qubits, one flag qubit, and two flag qubits, respectively. The probabilities are evaluated by considering two logical qubit states, $\ket{0}_L$ and $\ket{1}_L$. The case in which the number of detection events is zero implies that an error detected by the syndrome does not exist. In the case of $[3,1,3]_{f=0}$, the number of detection events to be zero is $62.325\%$ in the data, which implies that more than half become error-free. As a flag qubit is added, its probability of being zero decreases ($23.904\%$), and the value is the lowest when double-flag qubits are added ($0.322\%$) for distance three, which is the case of $[3, 1, 3]_{f=2}$. Further, the probability of the samples with an even number of detection events is more significant than that for the number of cases to be odd, regardless of distance and the number of flag qubits. This is because only the S error located at the boundary can produce an odd number of detection events. Notably, the probability of many detection events in the presence of flag qubits is more significant than that without any flag qubits.
\section*{Logical error rate}
 
Fig. \ref{fig11} depicts the logical error rates in the hardware-based detector graph (a, b, c) and the sample-based detector graph (d, e, f) with different numbers of flag qubits. The experiment is performed according to Fig. \ref{fig2}, and we obtain the average and standard deviation of the logical error rates for four different logical quantum states. The decoding results in the hardware-based and sample-based detector graphs exhibit similar behavior.
The logical error rate is obtained by comparing the initially prepared logical quantum state with the corrected state. Logical errors can occur even when the number of detection events is zero. Therefore, even when no detection event is detected, we should check whether a logical error exists. When several detection events are detected, the Pymatching algorithm is used to decode given detection events \cite{pymatching,pymatching2}. Finally, by applying a correction operator to the measured data qubit, the updated data qubit state can be a logical qubit state. If the corrected state is different from the initial state, we count a failure.\\

Fig. \ref{fig11} demonstrates the effectiveness of the syndrome extraction circuit with flag qubits in the repetition code on the IBM quantum machine. More specifically, in Fig. \ref{fig11} we can see that in \texttt{ibm\_kyoto}, the logical error rates of the repetition code exhibit a decrease as the distance of the repetition code increases from three to nine. Even though the average gate error rates for each structure increase as we consider more flag qubits, this tendency remains when flag qubits are present, regardless of whether the detector graph is hardware-based or sample-data-based. This implies that even when the data qubit is not adjacent to the syndrome qubit, a repetition code can operate on the IBM quantum computer. Moreover, even when there exist double-flag qubits between the data and syndrome qubits, a repetition code may operate on \texttt{ibm\_kyoto}.

\subsection*{Error suppression factor}
In this section, we present the error suppression factors $\Lambda$ for all cases, with and without flag qubits, based on fits to the experimental results shown in Fig. \ref{fig11}. The suppression factor ($\Lambda$), originally introduced in Ref. \cite{google_qec1}, quantitatively characterizes how rapidly the logical error rate decreases with increasing code distance.
\begin{equation}
  p_L = \frac{1-(1-2\epsilon_L)^R}{2},
  \label{eq:pL_fit}
\end{equation}
Let $R$ denote the number of syndrome extraction rounds, and let the logical error rate per round be approximated as $\epsilon_L\approx C/\Lambda^{(d+1)/2}$, where $C$ is a fitting constant, $\Lambda$ is the error suppression factor, and $d$ is the code distance. Then, the total logical error rate after $R$ rounds can be expressed as Equation \ref{eq:pL_fit}.

\begin{figure}[htbp!]
  \centering
  \includegraphics[width=\linewidth]{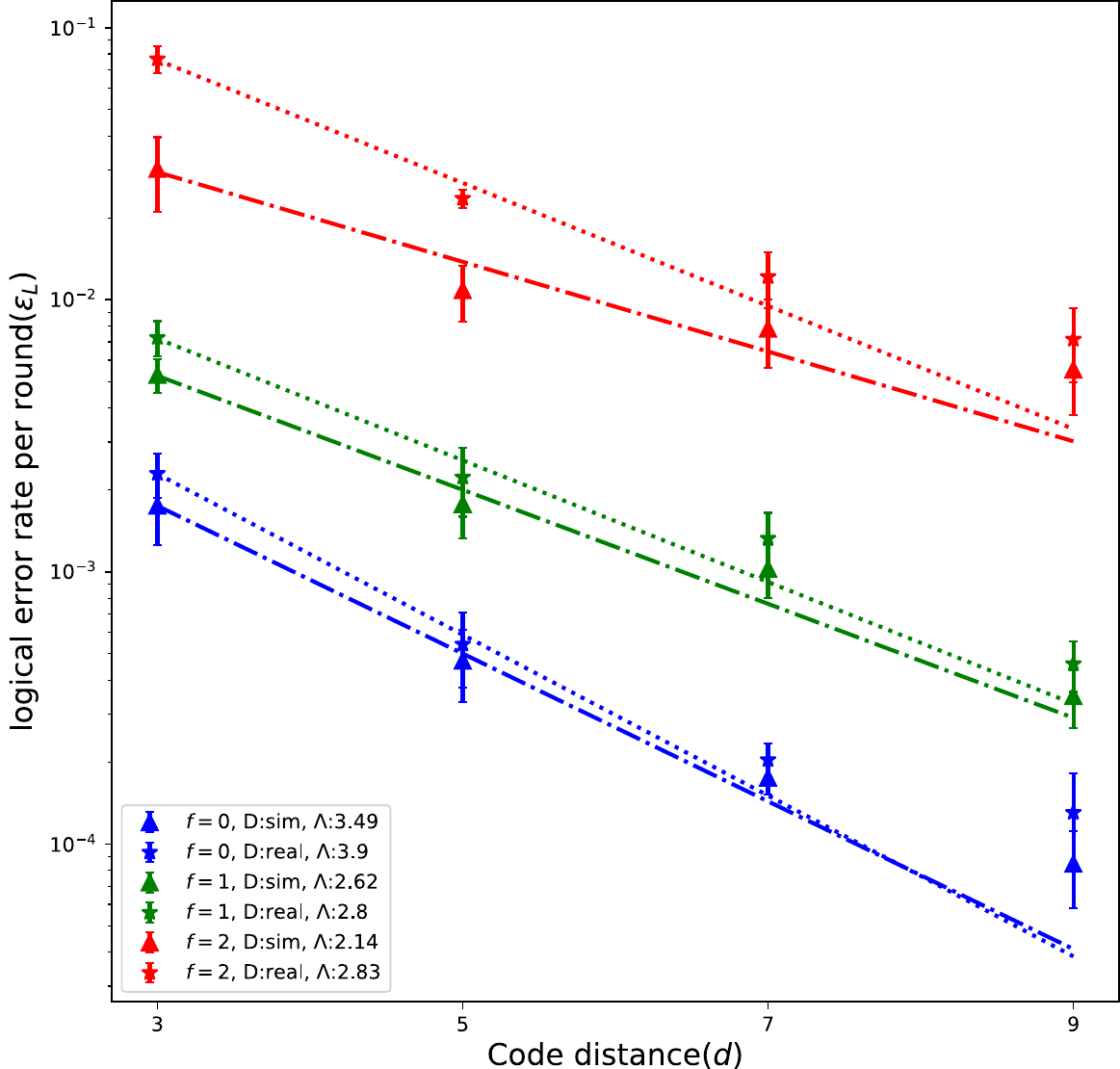}
  \caption{
  The logical error rate per round ($\epsilon_L$) is plotted as a function of the code distance ($d$) to plot the error suppression factor $\Lambda$ obtained through linear fitting. Error suppression factor $\Lambda$ for all cases, with and without flag qubits, where $f$ represents the number of flag qubits used between data and syndrome qubits. `D' denotes the decoding method, where \textit{sim} is hardware-based and \textit{real} is a sample-based one.
  }
  \label{fig:Lambda}
\end{figure}

Fig. \ref{fig:Lambda} shows the logical error rates per round as a function of the code distance. In the figure, the error suppression factors with or without flag qubits are represented by the slope of the fitted lines. The results obtained using the sample-based decoding method yield larger $\Lambda$ values compared to those from the hardware-based method. When no flag qubit is used with the hardware-based decoder, the suppression factor is $\Lambda \approx 3.49$. With the introduction of a single flag qubit and two flag qubits, the values decrease to $\Lambda \approx 2.62$ and $\Lambda \approx 2.14$, respectively. Notably, when using sampled data for decoding, the suppression factor $\Lambda$ increases regardless of the number of flag qubits. However, this difference diminishes as the code distance increases. It is worth noting that while the fitted line closely matches the logical error rates per round at small code distances, the data begins to deviate from the fitted line as the code distance increases. This deviation is particularly evident or higher code distances, such as $d=7$ or $d=9$, where logical error events become rare. Overall, increasing the code distance consistently suppresses logical error rates across all rounds of syndrome extraction, regardless of the presence of flag qubits, although introducing additional flag qubits tends to increase the logical error rates.

\section*{Conclusion}
 
In this study, we showed the effectiveness of a syndrome extraction circuit with flag qubits in the repetition code on the IBM quantum machine, \texttt{ibm\_kyoto}. {Quantum error correction was implemented using a repetition code with flag qubits was realized in the IBM quantum machine, \texttt{ibm\_kyoto}.} Because the IBM machine uses a heavy-hexagon structure, in which the maximum connection of a single node is three, the use of flag qubits may be required. We demonstrated that even when two flag qubits exist between a data qubit and a syndrome qubit, the logical error rates diminish as the distance between the repetition code increases from three to nine. This implies that a syndrome extraction circuit with flag qubits remains stable in the IBM machine for bit-flip or phase-flip errors.\\
 
{While flag qubits were originally introduced to detect a specific class of errors, namely hook errors that increase the effective error weight through qubit interactions, their utility in the repetition code may be limited by the code's inherent simplicity. In this context, incorporating flag qubits can even be counterproductive, as it adds physical qubits and elementary gates to the syndrome extraction circuit, thereby increasing the probability of logical errors. A compelling direction for future research is to determine how many flag qubits can be employed without compromising error suppression.\\}
 
In Ref. \cite{ibm_qec2,ibm_qec}, a hypergraph was constructed using the flag qubit information to correct errors. A promising direction for future research could be to explore the potential of the hypergraph in reducing the logical error rates in repetition codes with flag qubits. Moreover, increasing the number of flag qubits between the data and syndrome qubits is possible. Investigating how logical error rates vary with an increase in the number of flag qubits can be a simple experiment for assessing syndrome extraction circuits that employ long-range entanglement in situations where connectivity is limited.\\

We believe an intriguing future experiment would be to correct both bit- and phase-flip errors simultaneously and demonstrate error suppression with various numbers of flag qubits in the syndrome extraction circuit. The surface code can be a promising candidate for correcting both types of errors. However, unlike the repetition code, embedding the surface code on quantum processors with connectivity constraints, such as IBM quantum devices, is not a trivial problem. Suitable embedding schemes have been proposed in  Ref. \cite{strategy}. In contrast, incorporating flag qubits into the surface code could be simplified when using devices with a square lattice coupling map, for example, by placing flag qubits between data and syndrome qubits. While the number and placement of flag qubits cannot be arbitrarily chosen, it will be important to evaluate their effectiveness by testing different embedding schemes with varying numbers of flag qubits. In particular, this includes investigating how flag qubits can aid in error decoding by tracking error propagation in experiments and determining how many flag qubits can be used to maintain error suppression while scaling the code distance.\\

\section*{Acknowledgements}
This work was supported by the Basic Science Research Program through the National Research Foundation of Korea (NRF) funded by the Ministry of Education, Science, and Technology (NRF2022R1F1A1064459), and the Creation of the Quantum Information Science RD Ecosystem (Grant No. 2022M3H3A106307411) through the National Research Foundation of Korea (NRF) funded by the Korean government (Ministry of Science and ICT).

\section*{Data availability}
The datasets generated and analyzed for this study are available from the corresponding authors upon reasonable request.

\bibliographystyle{quantum} 
\bibliography{main}      

\begin{thebibliography}{10}

\bibitem{ibm_rep}
J.~R. Wootton and D.~Loss.
\newblock ``Repetition code of 15 qubits''.
\newblock \href{https://dx.doi.org/10.1103/PhysRevA.97.052313}{Physical Review
  A {\bf 97}, 052313}~(2018).

\bibitem{ibm_bell}
B~Hetényi and J.~R. Wootton.
\newblock ``Creating entangled logical qubits in the heavy-hex lattice with
  topological codes''.
\newblock \href{https://dx.doi.org/10.1103/PRXQuantum.5.040334}{PRX Quantum
  {\bf 5}, 040334}~(2024).

\bibitem{google_qec1}
{Google Quantum AI}.
\newblock ``Exponential suppression of bit or phase errors with cyclic error
  correction''.
\newblock \href{https://dx.doi.org/10.1038/s41586-021-03588-y}{Nature {\bf
  595}, 383}~(2021).

\bibitem{google_qec2}
{Google Quantum AI}.
\newblock ``Suppressing quantum errors by scaling a surface code logical
  qubit''.
\newblock \href{https://dx.doi.org/10.1038/s41586-022-05434-1}{Nature {\bf
  614}, 676--681}~(2023).

\bibitem{neutral}
D.~Bluvstein, S.~J. Evered, A.~A. Geim, S.~H. Li, H.~Zhou, T.~Manovitz, et~al.
\newblock ``Logical quantum processor based on reconfigurable atom arrays''.
\newblock \href{https://dx.doi.org/10.1038/s41586-023-06927-3}{NaturePages
  1--3}~(2023).

\bibitem{stabilizer}
A.~M. Steane.
\newblock ``Error correcting codes in quantum theory''.
\newblock \href{https://dx.doi.org/10.1103/PhysRevLett.77.793}{Phys. Rev. Lett.
  {\bf 77}, 793--797}~(1996).

\bibitem{gottesman}
D.~Gottesman.
\newblock ``Stabilizer codes and quantum error correction''.
\newblock \href{https://dx.doi.org/10.48550/arXiv.quant-ph/9705052}{PhD
  thesis}.
\newblock California Institute of Technology.
\newblock ~(1997).

\bibitem{boundary}
S.~B. Bravyi and A.~Y. Kitaev.
\newblock ``Quantum codes on a lattice with boundary''~(1998).
\newblock
  \href{http://arxiv.org/abs/quant-ph/9811052}{arXiv:quant-ph/9811052}.

\bibitem{topological}
E.~Dennis, A.~Kitaev, A.~Landahl, and J.~Preskill.
\newblock ``Topological quantum memory''.
\newblock \href{https://dx.doi.org/10.1063/1.1499754}{J. Math. Phys. {\bf 43},
  4452--4505}~(2002).

\bibitem{anyons}
A.~Y. Kitaev.
\newblock ``Fault-tolerant quantum computation by anyons''.
\newblock \href{https://dx.doi.org/10.1016/S0003-4916(02)00018-0}{Ann. Phys.
  {\bf 303}, 2--30}~(2003).

\bibitem{surface}
A.~G. Fowler, M.~Mariantoni, J.~M. Martinis, and A.~N. Cleland.
\newblock ``Surface codes: towards practical large-scale quantum computation''.
\newblock \href{https://dx.doi.org/10.1103/PhysRevA.86.032324}{Phys. Rev. A
  {\bf 86}, 032324}~(2012).

\bibitem{qecmemory}
B.~M. Terhal.
\newblock ``Quantum error correction for quantum memories''.
\newblock \href{https://dx.doi.org/10.1103/RevModPhys.87.307}{Rev. Mod. Phys.
  {\bf 87}, 307}~(2015).

\bibitem{FT1}
S.~Bravyi and A.~Kitaev.
\newblock ``Universal quantum computation with ideal clifford gates and noisy
  ancillas''.
\newblock \href{https://dx.doi.org/10.1103/PhysRevA.71.022316}{Phys. Rev. A
  {\bf 71}, 022316}~(2005).

\bibitem{FT2}
C.~Chamberland, P.~Iyer, and D.~Poulin.
\newblock ``Fault-tolerant quantum computing in the pauli or clifford frame
  with slow error diagnostics''.
\newblock \href{https://dx.doi.org/10.22331/q-2018-01-04-43}{Quantum {\bf 2},
  43}~(2018).

\bibitem{FT3}
D.~P. DiVincenzo and P.~Aliferis.
\newblock ``Effective fault-tolerant quantum computation with slow
  measurements''.
\newblock \href{https://dx.doi.org/10.1103/PhysRevLett.98.020501}{Phys. Rev.
  Lett. {\bf 98}, 020501}~(2007).

\bibitem{heavy-hexagon}
Y.~Kim et~al.
\newblock ``Evidence for the utility of quantum computing before fault
  tolerance''.
\newblock \href{https://dx.doi.org/10.1038/s41586-023-06096-3}{Nature {\bf
  618}, 500--505}~(2023).

\bibitem{hardware}
J.~B. Hertzberg et~al.
\newblock ``Laser-annealing josephson junctions for yielding scaled-up
  superconducting quantum processors''.
\newblock \href{https://dx.doi.org/10.1038/s41534-021-00464-5}{npj Quantum Inf.
  {\bf 7}, 1}~(2021).

\bibitem{flag}
R.~Chao and B.~W. Reichardt.
\newblock ``Fault-tolerant quantum computation with few qubits''.
\newblock \href{https://dx.doi.org/10.1038/s41534-018-0085-z}{npj Quantum
  Information{\bf 4}}~(2018).

\bibitem{flag2}
R.~Chao and B.~W. Reichardt.
\newblock ``Quantum error correction with only two extra qubits''.
\newblock \href{https://dx.doi.org/10.1103/PhysRevLett.121.050502}{Phys. Rev.
  Lett. {\bf 121}, 050502}~(2018).

\bibitem{flag3}
R.~Chao and B.~W. Reichardt.
\newblock ``Flag fault-tolerant error correction for any stabilizer code''.
\newblock \href{https://dx.doi.org/10.1103/PRXQuantum.1.010302}{PRX Quantum
  {\bf 1}, 010302}~(2020).

\bibitem{flag4}
C.~Chamberland and M.~E. Beverland.
\newblock ``Flag fault-tolerant error correction with arbitrary distance
  codes''.
\newblock \href{https://dx.doi.org/10.22331/q-2018-02-08-53}{Quantum {\bf 2},
  53}~(2018).

\bibitem{hh-code}
Y.~Kim, J.~Kang, and Y.~Kwon.
\newblock ``Design of quantum error correcting code for biased error on
  heavy-hexagon structure''.
\newblock \href{https://dx.doi.org/10.1007/s11128-023-03979-2}{Quantum
  Information Processing {\bf 22}, 230}~(2023).

\bibitem{magic}
Riddhi~S. Gupta et~al.
\newblock ``Encoding a magic state with beyond break-even fidelity''.
\newblock \href{https://dx.doi.org/10.1038/s41586-023-06846-3}{Nature {\bf
  625}, 259--263}~(2024).

\bibitem{strategy}
César Benito et~al.
\newblock ``Comparative study of quantum error correction strategies for the
  heavy-hexagonal lattice''~(2024).
\newblock  \href{http://arxiv.org/abs/2402.02185}{arXiv:2402.02185}.

\bibitem{d2_qec}
Christian~Kraglund Andersen et~al.
\newblock ``Repeated quantum error detection in a surface code''.
\newblock \href{https://dx.doi.org/10.1038/s41567-020-0920-y}{Nature Physics
  {\bf 16}, 875--880}~(2020).

\bibitem{d3_bacon}
Laird Egan et~al.
\newblock ``Fault-tolerant control of an error-corrected qubit''.
\newblock \href{https://dx.doi.org/10.1038/s41586-021-03928-y}{Nature {\bf
  598}, 281--286}~(2021).

\bibitem{d3_ion}
Ciaran Ryan-Anderson et~al.
\newblock ``Realization of real-time fault-tolerant quantum error correction''.
\newblock \href{https://dx.doi.org/10.1103/PhysRevX.11.041058}{Phys. Rev. X.
  {\bf 11}, 041058}~(2021).

\bibitem{d3_surf}
Sebastian Krinner et~al.
\newblock ``Realizing repeated quantum error correction in a distance-three
  surface code''.
\newblock \href{https://dx.doi.org/10.1038/s41586-022-04566-8}{Nature {\bf
  605}, 669--674}~(2022).

\bibitem{d3_dia}
M.~H. Abobeih et~al.
\newblock ``Fault-tolerant operation of a logical qubit in a diamond quantum
  processor''.
\newblock \href{https://dx.doi.org/10.1038/s41586-022-04819-6}{Nature {\bf
  606}, 884--889}~(2022).

\bibitem{d3_surper}
Youwei Zhao et~al.
\newblock ``Realization of an error-correcting surface code with
  superconducting qubits''.
\newblock \href{https://dx.doi.org/10.1103/PhysRevLett.129.030501}{Phys. Rev.
  Lett. {\bf 129}, 030501}~(2022).

\bibitem{ibm_qec}
N.~Sundaresan, T.~J. Yoder, Y.~Kim, et~al.
\newblock ``Demonstrating multi-round subsystem quantum error correction using
  matching and maximum likelihood decoders''.
\newblock \href{https://dx.doi.org/10.1038/s41467-023-38247-5}{Nat. Commun.
  {\bf 14}, 2852}~(2023).

\bibitem{melbourne}
S.~Gicev, L.~C. Hollenberg, and M.~Usman.
\newblock ``Quantum computer error structure probed by quantum error correction
  syndrome measurements''~(2023).
\newblock  \href{http://arxiv.org/abs/2310.12448}{arXiv:2310.12448}.

\bibitem{ibm_qec2}
C.~Chamberland, G.~Zhu, T.~J. Yoder, J.~B. Hertzberg, and A.~W. Cross.
\newblock ``Topological and subsystem codes on low-degree graphs with flag
  qubits''.
\newblock \href{https://dx.doi.org/10.1103/PhysRevX.10.011022}{Phys. Rev. X
  {\bf 10}, 011022}~(2020).

\bibitem{ibm_qec3}
E.~H. Chen et~al.
\newblock ``Calibrated decoders for experimental quantum error correction''.
\newblock \href{https://dx.doi.org/10.1103/PhysRevLett.128.110504}{Phys. Rev.
  Lett. {\bf 128}, 110504}~(2022).

\bibitem{qiskit}
{IBM Quantum and Community}.
\newblock ``Qiskit: An open-source framework for quantum computing''.
\newblock \url{https://qiskit.org}~(2021).

\bibitem{dd}
C.~Ryan-Anderson et~al.
\newblock ``Realization of real-time fault-tolerant quantum error correction''.
\newblock \href{https://dx.doi.org/10.1103/PhysRevX.11.041058}{Phys. Rev. X
  {\bf 11}, 041058}~(2021).

\bibitem{pymatching2}
O.~Higgott and C.~Gidney.
\newblock ``Sparse blossom: correcting a million errors per core second with
  minimum-weight matching''~(2023).
\newblock  \href{http://arxiv.org/abs/2303.15933}{arXiv:2303.15933}.

\bibitem{pymatching}
O.~Higgott.
\newblock ``Pymatching: A python package for decoding quantum codes with
  minimum-weight perfect matching''.
\newblock \href{https://dx.doi.org/10.1145/3505637}{ACM Transactions on Quantum
  Computing{\bf 3}}~(2022).

\bibitem{ecr}
N.~Sundaresan et~al.
\newblock ``Reducing unitary and spectator errors in cross resonance with
  optimized rotary echoes''.
\newblock \href{https://dx.doi.org/10.1103/PRXQuantum.1.020318}{PRX Quantum
  {\bf 1}, 020318}~(2020).

\bibitem{clifford1}
D.~Gottesman.
\newblock ``The heisenberg representation of quantum computers''~(1998).
\newblock
  \href{http://arxiv.org/abs/quant-ph/9807006}{arXiv:quant-ph/9807006}.

\bibitem{clifford2}
S.~Anders and H.~J. Briegel.
\newblock ``Fast simulation of stabilizer circuits using a graph-state
  representation''.
\newblock \href{https://dx.doi.org/10.1103/PhysRevA.73.022334}{Phys. Rev. A
  {\bf 73}, 022334}~(2006).

\bibitem{stim}
C.~Gidney.
\newblock ``Stim: a fast stabilizer circuit simulator''.
\newblock \href{https://dx.doi.org/10.22331/q-2021-07-06-497}{Quantum {\bf 5},
  497}~(2021).

\bibitem{virtual_Z}
D.~C. McKay, C.~J. Wood, S.~Sheldon, J.~M. Chow, and J.~M. Gambetta.
\newblock ``Efficient z gates for quantum computing''.
\newblock \href{https://dx.doi.org/10.1103/PhysRevA.96.022330}{Phys. Rev. A
  {\bf 96}, 022330}~(2017).

\bibitem{rd}
E.~Magesan, J.~M. Gambetta, and J.~Emerson.
\newblock ``Characterizing quantum gates via randomized benchmarking''.
\newblock \href{https://dx.doi.org/10.1103/PhysRevA.85.042311}{Phys. Rev. A.
  {\bf 85}, 042311}~(2012).

\end{thebibliography}
\clearpage

\section*{Supplementary information}

\subsection{Quantum hardware}

\subsubsection{Calibration table}

\begin{table}[htbp]
\centering

\begin{tabular}[t]{|| c | c | c ||}
    \hline\hline 
    \textbf{T1($\mu$s)} & \textbf{T2($\mu$s)} & \textbf{Readout (ns)} \\
    \hline
    217.69 & 140.21 & 1400 \\
    \hline\hline
    \multicolumn{2}{||c|}{\textbf{1Q gate time(ns)}} & \textbf{2Q gate time(ns)} \\
    \hline
    \multicolumn{2}{||c|}{60} & 660 \\
    \hline\hline
\end{tabular}
\quad 
\begin{tabular}[t]{|| c | c ||}
    \hline\hline
    \textbf{Readout} & \textbf{Idling} \\
    \hline
    $1.90 \times 10^{-2}$ & $2.79 \times 10^{-4}$ \\
    \hline\hline
    \textbf{1Q gate} & \textbf{2Q gate} \\
    \hline
    $2.79 \times 10^{-4}$ & $8.16 \times 10^{-3}$ \\
    \hline\hline
\end{tabular}
\caption{Time and error information for the physical qubits in $[9,1,9]_{f=2}$. The table lists the average times of T1, T2, readout, single-qubit gates, and two-qubit gates. The average error rates of readout, idling, single-qubit gates, and two-qubit gates are also displayed.}

\end{table}


Here, we describe the hardware specifications of \texttt{ibm\_kyoto}. The hardware has a heavy-hexagon structure consisting of transmons of a fixed frequency. \texttt{ibm\_kyoto} performs periodic calibrations in qubits and gates. Our study is based on the hardware specification at 
\parsepdfdatetime D:20231225053945+00'00'\endparsepdfdatetime
\theyear-\themonth-\theday-\thehour:\theminute:\thesecond\ $ \sim$
\parsepdfdatetime D:20231225114234+00'00'\endparsepdfdatetime
\thehour:\theminute:\thesecond\ \thetimezone. Values can vary over time, even within the period. Table 3 displays a summary of the hardware specifications at the time when we run and obtain data for $\ket{-}_L$ in $\lbrack9,1,9\rbrack _{f=2}$.

\begin{figure}
  \centering
  \includegraphics[width=0.5\textwidth]{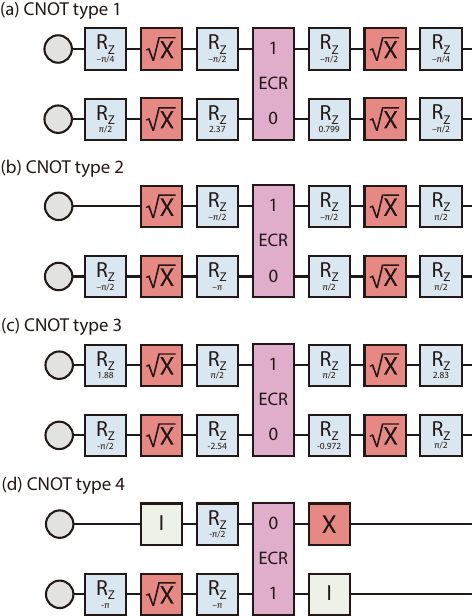}
  \caption{ Four different methods to implement the CNOT gate. The control and target qubits correspond to the upper and lower gray dots. The CNOT gate can be fragmented in four ways using the basis gate set from \texttt{ibm\_kyoto}. The structure varies depending on what parameters are used in the $R_Z$ gate and the type of the ECR gate.
  }\label{fig12}
  \end{figure}

The average values of time T1 and T2 of 49 physical qubits used are 217.69 $\mu$s and 140.21 $\mu$s, respectively. The longest operation among the hardware gates is the measurement, and its duration is 1400 ns. The 1Q gate is a single-qubit gate that contains $X$ and $\sqrt{X}$ except for $R_Z, I$. Because a virtual Z operation operates the $R_Z$ gate, we do not physically apply the gate during the execution of the quantum circuits. The duration of the single-qubit gate is 60 ns. Meanwhile, the duration of the two-qubit gate (ECR) is 660 ns. The average error rate of 1Q gate is $2.79 \times 10^{-4}$. The $I$-gate error is an idling error, a decoherence error caused by the free evolution of a single qubit. The error rate of the 1Q gate is the same as that of the idling error.

\subsubsection{ECR pulse}


The 2Q gate denotes a two-qubit gate obtained by the ECR pulse. There are two ways to operate the ECR gate: 1. $\mathrm{ECR}01={\frac{1}{\sqrt{2}}} (IX-XY)$; 2. $\mathrm{ECR}10 = {\frac{1}{\sqrt{2}}} (XI-YX)$. When the CNOT pulse comprises an ECR pulse and single-qubit gates, the sequences of the gates differ according to ECR01 or ECR10. Fig. \ref{fig12} depicts the four cases created by combining the ECR pulse and single-qubit gates. The physical CNOT gate is designed to apply an X gate to one physical qubit, known as the target qubit, based on the state of another physical qubit, which serves as the control qubit. In the Figure, each case shows the CNOT gate designed to implement two physical qubits, grey dots, by choosing the top node as the control qubit and the bottom node as the target qubit. Since idling errors occur because free evolution occurs before and after the ECR pulse, the depth of the circuit is identical to those in the other cases. Therefore, the common feature of these four cases is that we require two single-qubit gates before and after the ECR pulse. We adopt these cases to construct a noisy CNOT gate for a hardware-based decoder. A CZ gate can also be constructed in many ways by choosing different gate sets, such as the type of ECR pulse and parameters for single-qubit gates. However, the depth of the gates in the CZ gate is identical to that in the CNOT gate, and both have similar error channels. At the hardware level, these gate sequences can differ from what we apply physical pulses in a quantum circuit when optimizing them. However, we consider the logic gates where their noisy version is the worst case when modeling the error channel using its hardware calibration value.

\subsubsection{Circuit-level noise model}

\begin{figure}[t]
\centering
\includegraphics[width=\linewidth]{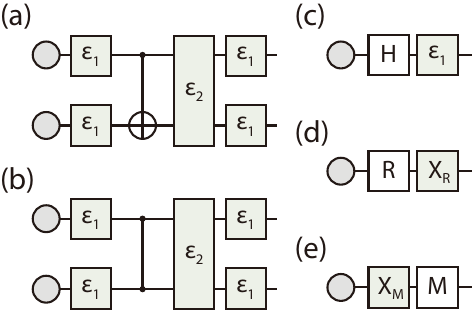}
\caption{ Error model for five elements of a syndrome extraction circuit: (a) CNOT, (b) CZ, (c) Hadamard, (d) reset, and (e) measurement gate. $\epsilon_1$ and $\epsilon_2$ denote the depolarized error channels of single-qubit and two-qubit gates, respectively. $X_R$ and $X_M$ correspond to the X error for the reset and measurement gate. The error channel surrounds the ideal gate to construct the noisy one. }\label{fig13}
\end{figure}


Every gate of the repetition code consists of a Clifford gate. The depolarizing error channel for the gates can be constructed, as depicted in Fig. \ref{fig13}, based on the conversion from the basis gate of the real device. Fig. \ref{fig13} (a), (b), (c), (d), and (e) depict the error models of the CNOT, CZ, Hadamard, reset, and measurement gates for constructing the hardware-based decoder with the Stim code. While X error channels are considered for reset and measurement gates, the depolarizing error channels are used for the other gates (CNOT, CZ, and H). We select the error rates for the error channels of the single- and two-qubit qubits based on their gate error rate in the actual quantum device by considering the qubits where the gates were applied. We emphasize the consideration of depolarizing error channels on physical qubits when inactive during the execution of physical hardware. Each syndrome extraction circuit comprises nine steps, including syndrome and flag qubit measurement. Certain physical qubits do not undergo quantum gates during each step and experience free evolution, making them susceptible to errors. When designing the hardware-based detector graph, depolarizing error channels are accounted for, considering their idling error rates from the calibration table.
 
\subsection{Repetition code on \texttt{ibm\_kyoto}}
\subsubsection{X syndrome extraction circuit}

\begin{figure}[t]
\centering
\includegraphics[width=\linewidth]{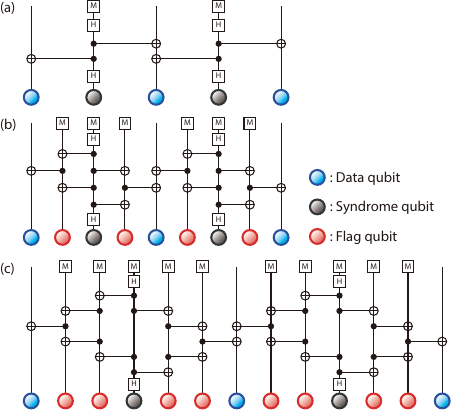}
\caption{ Quantum circuits for the X syndrome extraction without and with flag qubits. The quantum circuits for the X syndrome extraction round are shown when there is (a) no flag qubit, (b) a single-flag qubit, and (c) double-flag qubits between a data qubit and a syndrome qubit.}\label{fig14}
\end{figure}

The X syndrome extraction circuit detects the Z errors of the data qubits. Fig. \ref{fig14} depicts the X syndrome extraction circuit for the structures without a flag qubit and with a single-flag qubit or double-flag qubits between the data and syndrome qubits. When flag qubits exist, the construction of the quantum circuit of the X syndrome extraction circuit is identical to that of the Z stabilizer, except when using the CNOT operator and not the CZ operator to interact with data and its neighbor flag qubit. Initialization errors on either flag or syndrome qubits can be detected in the same way as we have covered. However, the error that affects the state of a logical qubit is not an X error but a Z error. Since the code considers only Z errors and all data qubits are initialized on an X basis, the state of the logical qubit we are interested in is unaffected by X errors.

\subsubsection{Selected qubit lists}
\begin{figure*}[t]
\centering
\includegraphics[width=\linewidth]{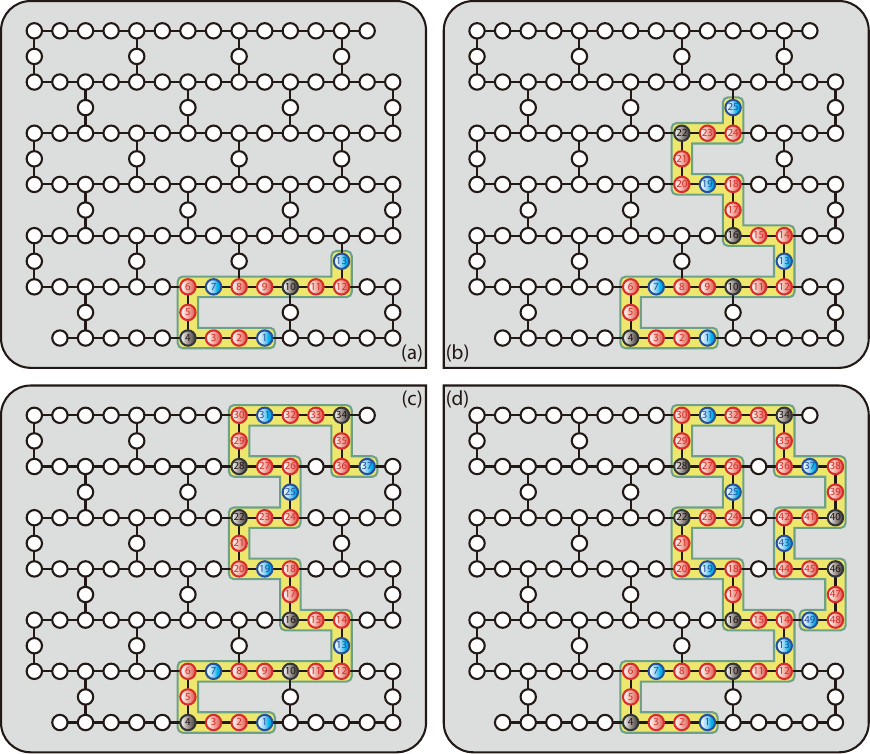}
\caption{ Selected physical qubits for different distances $d=3, 5, 7, 9$. The physical qubits that used in
(a) $\lbrack3,1,3\rbrack _{f=2}$, 
(b) $\lbrack5,1,5\rbrack _{f=2}$,
(c) $\lbrack7,1,7\rbrack _{f=2}$, and 
(d) $\lbrack9,1,9\rbrack _{f=2}$ are shown.
The blue, red, and black dots correspond to data, flag, and syndrome qubits.
}\label{fig15}
\end{figure*}

\begin{figure*}[t]
\centering
\includegraphics[width=\linewidth]{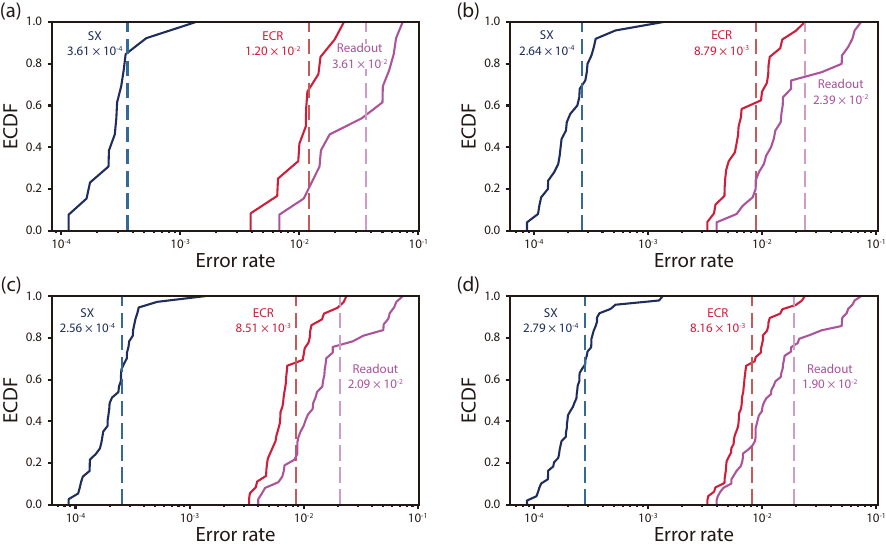}
\caption{ Gate error rates of the physical qubits in the repetition codes with double-flag qubits. The figure shows the case of
(a) $\lbrack3,1,3\rbrack _{f=2}$,
(b) $\lbrack5,1,5\rbrack _{f=2}$,
(c) $\lbrack7,1,7\rbrack _{f=2}$, and
(d) $\lbrack9,1,9\rbrack _{f=2}$. The ECDF is plotted as the function of the error rates of the basis gate set of \texttt{ibm\_kyoto} for each case.}\label{fig16}
\end{figure*}
\begin{figure*}[t]
\centering
\includegraphics[width=\linewidth]{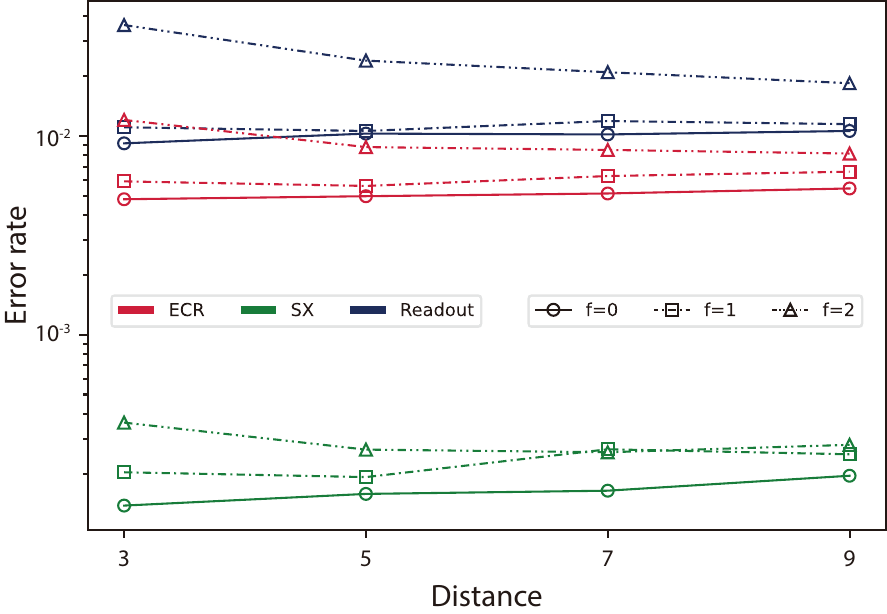}
\caption{ Average error rates for each gate with different structures. The red, green, and blue markers represent the average error rate of ECR, single qubit gate, and readout, respectively. Meanwhile, the solid, dash-dotted, and dash double-dotted lines denote the structure with no flag qubit, a single-flag qubit, and double-flag qubits. As we add more flag qubits, the average error rates of the gates increase. }\label{fig17}
\end{figure*}
 We select the physical qubits for the double-flag qubit structure, as depicted in Fig. \ref{fig15}. When the logical qubit is initialized as $|-\rangle_L$, the average error rate for the error type for each case is as follows. The average readout error of $[3,1,3]_{f=0}$($[9,1,9]_{f=0}$) is $9.18 \times 10^{-3}$($1.06 \times 10^{-2}$). The average readout error of $[3,1,3]_{f=1}$($[9,1,9]_{f=1}$) is $1.11 \times 10^{-2}$($1.15 \times 10^{-2}$). Meanwhile, the average error rate of the ECR gate of $[3,1,3]_{f=0}$($[9,1,9]_{f=0}$) is $4.80 \times 10^{-3}$($5.44 \times 10^{-3}$). The average error rate of the ECR gate of $[3,1,3]_{f=1}$($[9,1,9]_{f=1}$) is $5.91 \times 10^{-3}$($6.61 \times 10^{-3}$). Fig. \ref{fig16} depicts the qubits and error rates of gates when two flag qubits are introduced between a data qubit and a syndrome qubit. \texttt{ibm\_kyoto} provides 127 qubits. Therefore, the qubit selection for the syndrome extraction circuit with flag qubits in the repetition code is not unique. We use the function ``transpile" from the qiskit package to choose specific physical qubits that are likely to provide the best performance among the possible combinations. In Fig. \ref{fig17}, the average error rates for each basis gate of the selected physical qubits in the hardware are plotted as a function of the distance and number of flag qubits between the data and syndrome qubits. As more flag qubits are added to the structure, the average error rates for all gates increase.
 
\subsubsection{Syndrome calculation}

\begin{figure}[htbp!]
\centering
\includegraphics[width=\linewidth]{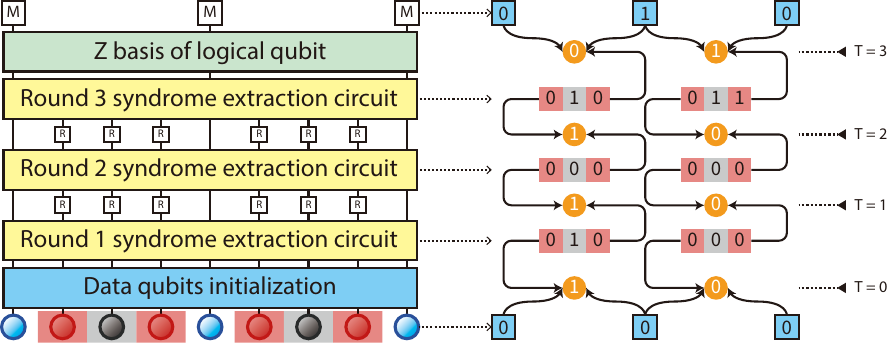}
\caption{ Procedure for extracting the syndrome from the sampled data. The structure is $[3,1,3]_{f=1}$ and the number of the syndrome extraction circuit rounds is three. The initial logical qubit state is $|0\rangle_L$. Each syndrome extraction circuit produces six outcomes from the flag and syndrome qubits. Their outcomes can be grouped with red and black boxes. The syndrome bits corresponding to the orange dots can be calculated by checking the parity of two consecutive groups in time. }\label{fig18}

\end{figure}

Here, we explain the method used to obtain the syndromes from the outcomes of syndrome extraction circuits. When R round syndrome extraction is performed, the syndrome can be broadly divided into three parts in terms of time $T$: 1. at $T = 0$, 2. from $T = 1$ to $T = (R-1)$, and 3. at $T = R$. The method used to treat this syndrome is depicted in Fig. \ref{fig18}. The diagram depicts an example of a method for obtaining a syndrome with $[3,1,3]_{f=1}$ and $R = 3$. The initial data qubits are prepared with $|0\rangle_L = |000\rangle$. The outcomes from the syndrome extraction circuits can be divided into patches based on the syndrome qubits, and each patch is covered with red and black boxes containing three bits from the syndrome and flag qubits in Fig. \ref{fig18}. The calculation of a syndrome bit involves checking the parity of six values between two consecutive patches. The process is as follows: the system counts the number of occurrences of the digit `1' among these six values. If the count is odd, the syndrome bit is set to `1' otherwise, it is set to `0'. This method determines the syndrome bit based on the parity of observed values within a specified temporal range. For a syndrome bit located in the temporal boundary, two data qubits that are neighbors of a patch are used to calculate the syndrome bit. For instance, each syndrome bit in the initial syndrome round at T = 0 utilizes five bits comprising three values from the outcomes of the first round of a syndrome extraction circuit and two values from the initial states of its adjacent data qubits. The process is the same for the syndrome bits in the final rounds ($T = 3$), except that the measured states of the data qubits are used instead of their initial states. The method for calculating the syndrome bit-string can be applied similarly to a structure involving double-flag qubits. The only variation lies in the specific number of bits, which is five in the case of double-flag qubits between the data and syndrome qubits within each patch.

\subsubsection{Correlation matrix}

\begin{figure}[htbp!]
\centering
\includegraphics[width=\linewidth]{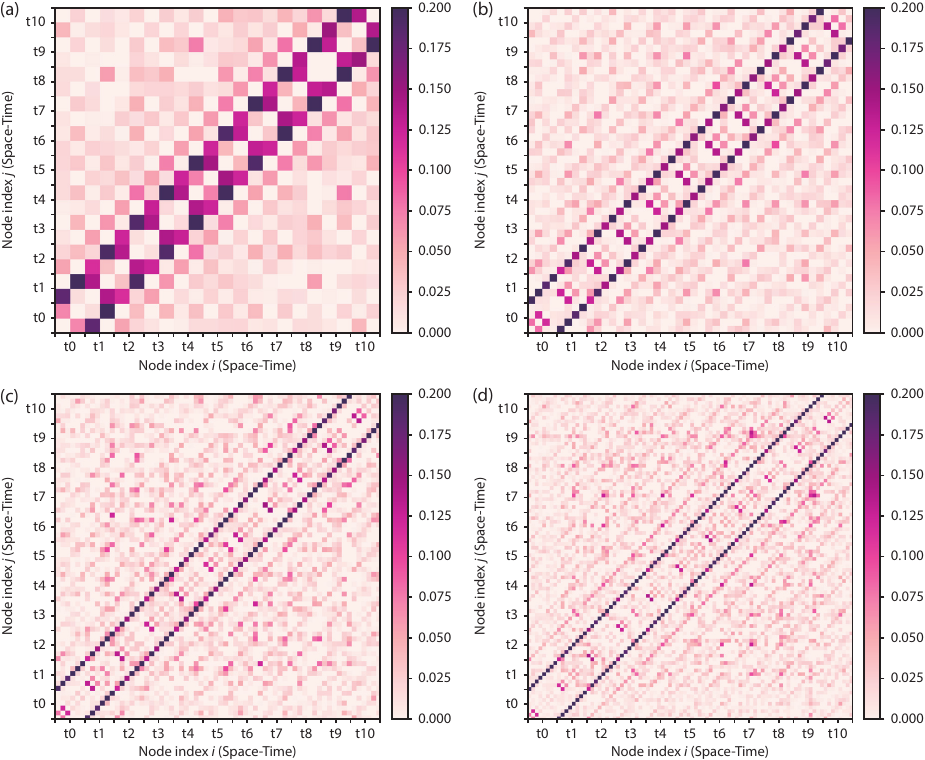}
\caption{ Space-Time correlation matrices obtained in the structures with double-flag qubits. Space-Time correlation matrices are shown in the case of (a) $[3,1,3]_{f=2}$, (b) $[5,1,5]_{f=2}$, (c) $[7,1,7]_{f=2}$, and (d) $[9,1,9]_{f=2}$. The data is collected from the experiment with $|-\rangle_L$ state. Ten rounds of the syndrome extraction circuit are implemented. The number of samples is 50,000.}\label{fig19}

\end{figure}
\begin{figure}[htbp]
\centering
\includegraphics[width=\linewidth]{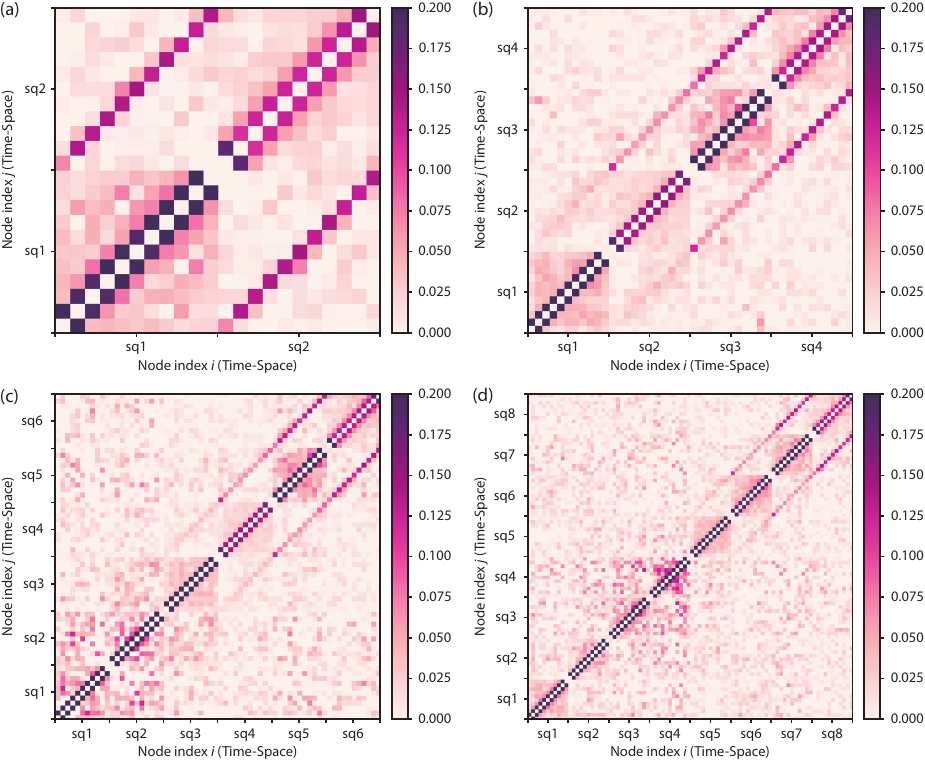}
\caption{ Time-Space correlation matrices obtained in the structures with double-flag qubits.  Time-Space correlation matrices are shown in the case of (a) $[3,1,3]_{f=2}$, (b) $[5,1,5]_{f=2}$, (c) $[7,1,7]_{f=2}$, and (d) $[9,1,9]_{f=2}$. The data is collected from the experiment with $|-\rangle_L$ state. Ten rounds of the syndrome extraction circuit are implemented. The number of samples is 50,000.
    }\label{fig20}
\end{figure}

The correlation matrix expresses the correlation between two detectors in a detector graph. Fig. \ref{fig19} and \ref{fig20} depict the correlation matrix obtained from 10 rounds of the X syndrome extraction circuit when $\ket{-}_L$ is prepared, and the number of samples is 50,000 in the case of double-flag qubits. The matrices are displayed with the space-time and time-space methods depicted in Fig. \ref{fig19} and \ref{fig20}, respectively. The numbering in space-time is performed in terms of space, whereas that in time-space is processed in terms of time. Each pixel in the correlation matrix corresponds to an error probability, and the representation excludes negative values. 
 Moreover, the color scale is truncated to enhance visibility and clarity, with 0.2 as the maximum value. We emphasize positive error probabilities within the specified color scale while disregarding negative values to represent the error patterns accurately. The T errors are dominant for all cases, consistent with the hardware specifications. 
 
 In Ref \cite{google_qec1}, the cross-talk effects, observable in Time-Space correlation matrices, happen due to physical qubits locally nearby at the hardware level, even though they are not physically interacting in a quantum circuit. More specifically, the cross-talk effects can be seen from the correlation between syndrome bits that are spatially close in the real device but not neighbors in the perspective of the code. However, we are unable to observe these effects. In cases where we run the repetition code without flag qubits, this could be due to the initially selected physical qubits being locally separated at the hardware level. When flag qubits are employed in the syndrome extraction circuits, this lack of observation may be attributed to relatively high error rates resulting from including flag qubits in the structure, diminishing the visibility of these effects.
 
 \subsubsection{Detection event rate}

\begin{figure}[htbp]
\centering
\includegraphics[width=\linewidth]{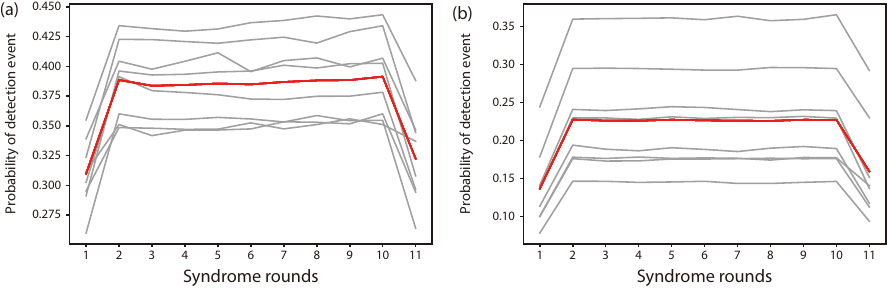}
\caption{ The probability of detection events. The probability of detection events for the $\ket{-}_L$ state in the $[9,1,9]_{f=2}$ structure is shown in the case of (a) real quantum device, \texttt{ibm\_kyoto}, and (b) Stim code. The data is collected from the experiment with $|-\rangle_L$ state. Here, the number of syndrome rounds is eleven. The number of samples is 50,000.}\label{fig21}
\end{figure}

We define the detection event rate as the ratio of the number of samples with detection events to the total number of samples. Fig. \ref{fig21} depicts the detection event rate(gray line) from each syndrome qubit for $[9,1,9]_{f=2}$ versus the syndrome rounds. The red line represents the average of all detection event rates for each syndrome qubits. The first (last) round also uses the initial (measured) state of the data qubits to identify the detection events. Data are obtained by initializing the logical qubit as $\ket{-}_{L}$ with 50,000 samples. The number of syndrome extraction cycles is 10. Thus, there are 11 rounds of syndromes. We use data sampled from (a) \texttt{ibm\_kyoto} and (b) the Stim code, which adopts the noise model of \texttt{ibm\_kyoto}. Both cases have similar behavior regarding the probability of detection events being lower in the first and last rounds than in other rounds and maintaining one value between them. However, those values are not the same, which can be attributed to the oversimplification of each qubit's error channel as a depolarizing and idling error model. This suggests that the system on the device is more vulnerable to errors than we initially modeled.

\subsubsection{Logical error rates}

\begin{figure}[htbp]
\centering
\includegraphics[width=\linewidth]{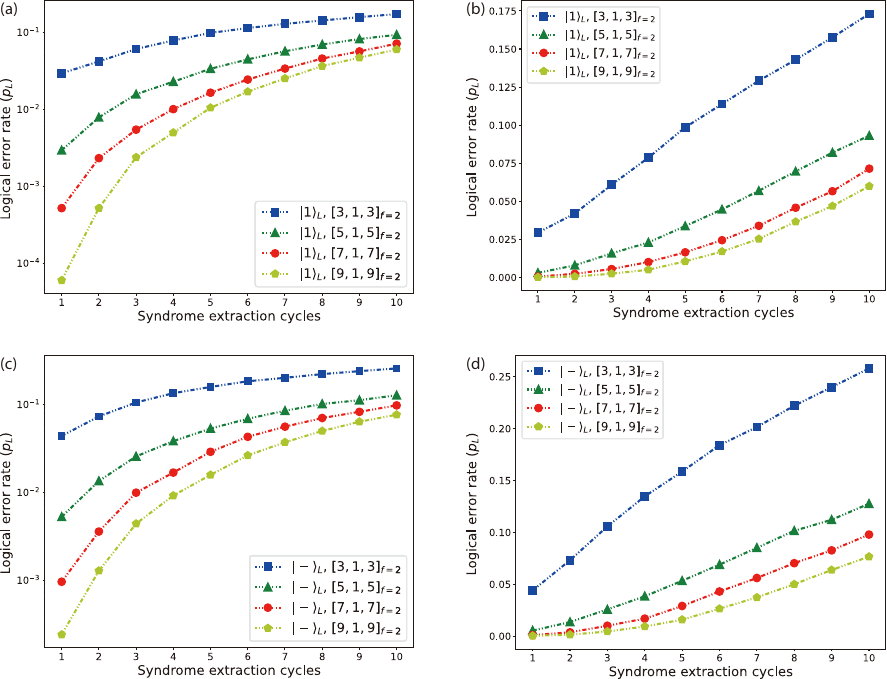}
\caption{ Logical error rates for the structure with double-flag qubits and different initial logical states.
The logical error rates in the case of $\ket{1}_L$ are plotted on (a) a log scale and (b) a linear scale for the vertical axis. Likewise, the logical error rates in the case of $\ket{-}_L$ are plotted on (c) a log scale and (d) a linear scale for the vertical axis. The logical error rates are obtained from the hardware-based decoder.
}\label{fig22}
\end{figure}

The logical error rates for each initial state are evident in Fig. \ref{fig22}. The Hardware-based decoder calculates the number of faults and corrected samples. To sample the data from \texttt{ibm\_kyoto}, the experimental implementation is the same, except for the initial state of the logical qubit and syndrome extraction circuit based on the basis: 1. The Z basis logical qubit ($\left|1\right\rangle_L$) 2. X basis logical qubit ($\left|-\right\rangle_L$). The Z(X) basis logical qubit uses a syndrome extraction circuit constructed by the Z(X) stabilizers and considers only X(Z) errors in the system.

\subsubsection{Samples}

\begin{figure}[htbp]
\centering
\includegraphics[width=\linewidth]{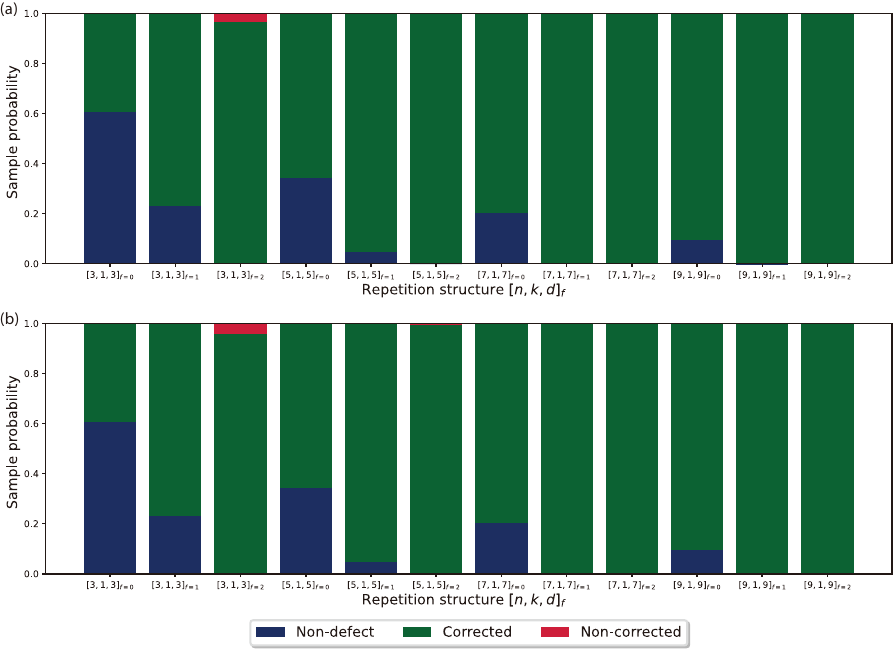}
\caption{ 
  The probability of the categorized results for all cases of repetition codes.
  The data from (a) the hardware-based and (b) sample-based decoder is categorized. The samples can be divided into three classes: non-defect(blue bar), corrected(green bar), and non-corrected(red bar). Each case with a different structure regarding the distance and the number of flag qubits is indicated using the stacked bars.}\label{fig23}
\end{figure}

 The sampled data are categorized into non-detected, corrected, and non-corrected. Non-detected samples are those in which the detection event is not detected due to the absence of error. Corrected samples correspond to the data in which detection events are detected and corrected; if these are not corrected, they belong to the non-corrected samples. In Fig. \ref{fig23}, the ratio of the categorized samples is shown with stacked bar charts for all cases of $[d,1,d]_{f}$ for both the hardware- and sample-based decoders. Although the samples are non-defective, a logical error can occur, which can be sorted as non-corrected instead of non-defective. As the structure size increases, the proportion of non-defects decreases. It is worth pointing out that most data samples are categorized as corrected samples as we increase the structure size.

\onecolumn
\clearpage
\clearpage
\clearpage
\clearpage

 \clearpage
 \clearpage
\end{document}